\documentclass[twocolumn, twocolappendix]{aastex631}

\usepackage{graphicx}	
\usepackage{amsmath}	
\usepackage{upgreek}    
\usepackage{multirow}   
\usepackage[normalem]{ulem} 
\usepackage[caption=false]{subfig}
\usepackage{xcolor}
\usepackage{CJKutf8} 
\usepackage{comment}
\usepackage[encapsulated]{CJK}
\newcommand{\cntext}[1]{\begin{CJK}{UTF8}{gbsn}#1\end{CJK}}

\newcommand{\sgra}{Sgr$\,$A*\,}

\received{May 9, 2023}
\accepted{August 28, 2023}

\graphicspath{./}

\begin{document}
\begin{CJK*}{UTF8}{gbsn} 

\title{A search for pulsars around \mbox{Sgr$\,$A*} in the first Event Horizon Telescope dataset}

\correspondingauthor{Pablo Torne}
\email{torne@iram.es}

\correspondingauthor{Kuo Liu}
\email{kliu@mpifr-bonn.mpg.de}

\author[0000-0001-8700-6058]{Pablo Torne}
\affiliation{Institut de Radioastronomie Millim\'etrique, Avenida Divina Pastora 7, Local 20, E-18012, Granada, Spain}
\affiliation{Max-Planck-Institut f\"ur Radioastronomie, Auf dem H\"ugel 69, D-53121 Bonn, Germany}

\author[0000-0002-2953-7376]{Kuo Liu}
\affiliation{Max-Planck-Institut f\"ur Radioastronomie, Auf dem H\"ugel 69, D-53121 Bonn, Germany}

\author[0000-0001-6196-4135]{Ralph~P.~Eatough}
\affiliation{National Astronomical Observatories, Chinese Academy of Sciences, 20A Datun Road, Chaoyang District, Beijing 100101, P. R. China}
\affiliation{Max-Planck-Institut f\"ur Radioastronomie, Auf dem H\"ugel 69, D-53121 Bonn, Germany}

\author[0000-0002-7730-4956]{Jompoj Wongphechauxsorn}
\affiliation{Max-Planck-Institut f\"ur Radioastronomie, Auf dem H\"ugel 69, D-53121 Bonn, Germany}

\author[0000-0002-6156-5617]{James M. Cordes}
\affiliation{Cornell Center for Astrophysics and Planetary Science, Cornell University, Ithaca, NY 14853, USA}

\author[0000-0003-3922-4055]{Gregory Desvignes}
\affiliation{Max-Planck-Institut f\"ur Radioastronomie, Auf dem H\"ugel 69, D-53121 Bonn, Germany}
\affiliation{LESIA, Observatoire de Paris, Université PSL, CNRS, Sorbonne Université, Université de Paris, 5 place Jules Janssen, 92195 Meudon, France}

\author[0000-0002-9945-682X]{Mariafelicia De Laurentis}
\affiliation{Dipartimento di Fisica ``E. Pancini'', Universit\'a di Napoli ``Federico II'', Compl. Univ. di Monte S. Angelo, Edificio G, Via Cinthia, I-80126, Napoli, Italy}
\affiliation{Institut f\"ur Theoretische Physik, Goethe-Universit\"at Frankfurt, Max-von-Laue-Stra{\ss}e 1, D-60438 Frankfurt am Main, Germany}
\affiliation{INFN Sez. di Napoli, Compl. Univ. di Monte S. Angelo, Edificio G, Via Cinthia, I-80126, Napoli, Italy}

\author[0000-0002-4175-2271]{Michael Kramer}
\affiliation{Max-Planck-Institut f\"ur Radioastronomie, Auf dem H\"ugel 69, D-53121 Bonn, Germany}
\affiliation{Jodrell Bank Centre for Astrophysics, School of Physics and Astronomy, The University of Manchester, Manchester M13 9PL, UK}

\author[0000-0001-5799-9714]{Scott M. Ransom}
\affiliation{National Radio Astronomy Observatory, 520 Edgemont Road, Charlottesville, VA 22903, USA}

\author[0000-0002-2878-1502]{Shami Chatterjee}
\affiliation{Cornell Center for Astrophysics and Planetary Science, Cornell University, Ithaca, NY 14853, USA}

% ...







\begin{abstract}

The Event Horizon Telescope (EHT) observed in 2017 the supermassive black hole at the center of the Milky Way, \mbox{Sagittarius$\,$A*} (\mbox{Sgr$\,$A*}), at a frequency of 228.1$\,$GHz ($\uplambda=1.3\,$mm). 
The fundamental physics tests that even a single pulsar orbiting \mbox{Sgr$\,$A*} would enable motivate searching for pulsars in EHT datasets.
The high observing frequency means that pulsars $-$ which typically exhibit steep emission spectra $-$ are expected to be very faint. However, it also negates pulse scattering, an effect that could hinder pulsar detections in the Galactic Center. 
Additionally, magnetars or a secondary inverse Compton emission could be stronger at millimeter wavelengths than at lower frequencies. We present a search for pulsars close to \mbox{Sgr$\,$A*} using the data from the three most-sensitive stations in the EHT 2017 campaign: the Atacama Large Millimeter/submillimeter Array, the Large Millimeter Telescope and the IRAM~$30\,{\rm m}$ Telescope. We apply three detection methods based on Fourier-domain analysis, the Fast-Folding-Algorithm and single pulse search targeting both pulsars and burst-like transient emission; using the simultaneity of the observations to confirm potential candidates.
No new pulsars or significant bursts were found. Being the first pulsar search ever carried out at such high radio frequencies, we detail our analysis methods and give a detailed estimation of the sensitivity of the search. We conclude that the EHT 2017 observations are only sensitive to a small fraction ($\lesssim2.2$\%) of the pulsars that may exist close to \mbox{Sgr$\,$A*}, motivating further searches for fainter pulsars in the region.

\end{abstract}




\section{Introduction} \label{sec:intro}
The first test of strong-field gravity came from measurements of the relativistic orbital decay in a binary pulsar system \citep{tw+82, tw+89},
with subsequent tests of increasing  precision using other binary (or even triple) pulsar systems~\citep[see e.g.][for a review]{wk+20}. So far, all gravity experiments using pulsars conform with the predictions of general relativity (GR), but it is expected that the most constraining tests will come from a pulsar black hole binary system \citep[][]{wk99,lewk+14}, and in particular from pulsars orbiting Sagittarius~A*~\mbox{(Sgr$\,$A*)} itself \citep[][]{kbc+04,lwk+12}. Discovery and timing observations of pulsars near \mbox{Sgr$\,$A*} could  provide measurements of the spin and quadrupole moment of the supermassive black hole (SMBH), in addition to unique information on Galactic Centre (GC) stellar populations, dark matter, and the $\gamma-$ray excess \citep{fermi-gc+16, day16, bar16},
along with measurements of the magnetoionic plasma 
\citep[][]{eat13,des18}.

Together with the groundbreaking measurements of binary black hole mergers with gravitational wave detectors \citep{Abbott+16} and high-precision astrometry of stars orbiting \mbox{Sgr$\,$A*} \citep{grvty+18,gravc19,grav20_precS2,Do2019},
millimeter interferometric imaging of SMBHs by the Event Horizon Telescope (EHT) Collaboration represents a transformation in the way black holes can be observed and studied 
\citep[see][and references therein for an overview of key results]{eht19a,eht22a}. Comparison of EHT image properties with synthetic images derived from general relativistic magnetohydrodynamic (GRMHD) simulations, and general relativistic ray tracing, provide a new framework in which to measure the fundamental properties of black holes and test theories of gravity in the strong-field regime \citep[e.g.,][]{myf+18,eht19e,eht19f,eht22f,ofp+22,you23}. 

EHT images of the SMBHs Messier~87* (M87*) and \mbox{Sgr$\,$A*} are consistent with predictions for the black hole shadow of a spinning Kerr black hole in GR, whilst certain alternative theories of gravity, which deviate from the Kerr metric in the strong-field regime, have been ruled out \citep{psaltis+20}. For \mbox{Sgr$\,$A*}, the EHT images can be synergized with the measurements from an orbiting pulsar. For example \citet{psal16} showed that the spin and quadrupole moment of \mbox{Sgr$\,$A*} from the motion of orbiting stars \citep[e.g.][]{2005ApJ...622..878W} and pulsars have correlated uncertainties that are almost orthogonal to those from black hole shadow images, thereby increasing overall measurement accuracy. 

While estimates of the GC pulsar population ranges from about a few to several thousands \citep[e.g.,][]{pl04, whar12, dexo14, rajwade2017, schoedel20, chen2023},
only six radio pulsars have been detected within $50\,{\rm pc}$ (projected) from \mbox{Sgr$\,$A*} \citep{john06,den09}, at cm wavelengths, including the radio magnetar PSR~J1745$-$2900 that lies just three arcseconds (i.e., a projected distance of $0.1\,{\rm pc}$) away \citep{ken13, eat13, rea13}.
Although more sensitive pulsar searches of the GC region have been conducted \citep[e.g.,][]{mq10, etd+21,acc+22}, they failed to discover additional pulsars, potentially highlighting the limits of surveys at cm wavelengths.

Multipath propagation by scattering from electron-density inhomogeneities typically broadens radio pulse emission, reducing the signal-to-noise ratio of periodic or single-burst radio signals \cite[][]{cl97}. The severity of this effect depends on the observing frequency $\nu$, ($\sim\nu^{n}$; where $n\approx -4$), spin period ($P$), pulse or burst width, and also on the scattering environment toward the GC. The ${\rm NE}2001$ model of the  Galactic distribution of free electrons \citep{cl2002} predicts a large scattering time scale that is inconsistent with the combined measurements of PSR~J1745$-$2900 and \mbox{Sgr$\,$A*} \citep{spi14,bow14,bdd+15} but may apply to other lines of sight toward $(\lesssim0.5^\circ)$ the GC \citep{dex2017GCscatt}. Therefore, one rationale is to search the GC at mm and sub-mm wavelengths where multipath scattering can be avoided altogether. However, given the average steep spectrum of pulsars' radio emission where flux density $\sim \nu^{\alpha}$, with $\alpha$ typically between $-1$ and $-2.5$
\citep[see e.g.,][]{jank18}, many pulsars are expected to be faint at frequencies where scattering is negligible. Nevertheless, a subset of pulsars and their high magnetic field counterparts, magnetars, show flat or inverted spectra\footnote{Around $13\%$ of known pulsars with a measured spectral index have $\alpha\ge-1.0$ \citep[][PSRCAT in 2023]{mhth05}.
} with detectable mm and sub-mm emission \citep{cam07b,tor20,chu21,tor22}, including from PSR~J1745$-$2900 itself \citep{tek+15,tde+17}.

Additional physical processes may contribute to emission in these bands and could enhance the probability of detection. For example, induced inverse Compton scattering of low-frequency radio photons at frequency $\nu$ can produce higher-frequency photons at $\nu^{\prime} \sim \gamma^2 \nu$ in some pulsar models \cite[e.g.][]{1976MNRAS.174...59B, philkra22}, where $\gamma$ is the Lorentz factor. For example, with $\nu = 100\,{\rm MHz}$ and $\gamma = 50$,  scattered photons will boost emission at $\nu \sim 250\,{\rm GHz}$. Since a wide range of Lorentz factors is possible, it is necessary to search in a corresponding wide range of observing bands with sensitive mm and sub-mm telescopes. 

The first attempts at searching the GC for pulsars at mm wavelengths were undertaken with the IRAM~$30\,{\rm m}$ telescope $(\uplambda \simeq 2,\,3\,{\rm mm})$ and the Atacama Large Millimeter/submillimeter Array $(\uplambda \simeq 3.5\,{\rm mm})$ \citep{tor21,lde+21}. In this work we report the first $1.3\,{\rm mm}$ searches for GC pulsars using data from the EHT 2017 campaign on \mbox{Sgr$\,$A*}. As well as utilizing the most sensitive mm/sub-mm telescopes, the simultaneous EHT observations from multiple telescope sites enable efficient rejection of false positive detections.

In Section~\ref{sec:obs} we describe the observations and data processing; in Section~\ref{sec:results} the results of our pulsar search of the EHT data are detailed; Section~\ref{sec:discussion} is a discussion of the findings; Section~\ref{sec:futureimp} presents potential improvements for the future and Section~\ref{sec:concs} summarizes our conclusions.

\section{Observations and data processing} \label{sec:obs}

\subsection{Observations} \label{ssec:obs}

The EHT 2017 observing campaign was scheduled on five nights during 2017 April 5–14, where three nights (April 6, 7 and 11) included exposures to \mbox{Sgr$\,$A*} ($\alpha_\mathrm{J2000} = 17^\mathrm{h}45^\mathrm{m}40^\mathrm{s}.0361$, $\delta_\mathrm{J2000} = -29^\mathrm{\circ}00\mathrm{'}28\mathrm{''}.168$). 
At each epoch, the track on \mbox{Sgr$\,$A*} was divided into individual scans of approximately 7$-$10\,min each, switching between \mbox{Sgr$\,$A*} and an active galactic nucleus calibrator source, J1924$-$2914 and/or NRAO$\,$530. Baseband voltage data were recorded with 2-bit sampling at a total rate of 32\,Gbps in two 2-GHz
IF-bands centered at 229.1 and 227.1\,GHz, labeled ``high'' and ``low'', respectively. 
More details of the EHT array, the observations and data recording can be found in \cite{eht19b,eht19c,eht22b}. 

Since system sensitivity is one of the primary considerations for a pulsar search of the GC, we chose to analyze data only from the three most sensitive telescopes in the EHT 2017 observations: the phased Atacama Large Millimeter/submillimeter Array (ALMA), the Large Millimeter Telescope Alfonso Serrano (LMT) and the IRAM~$30\,{\rm m}$ telescope (PV). While the sensitivity of phased ALMA is significantly higher than the other two stations, its field of view, given by the full width at half maximum (FWHM) of its synthetic beam during the EHT 2017 observations on \mbox{Sgr$\,$A*}, is only 1--2\arcsec \citep{gmm+21}; offering comparatively limited sky coverage. LMT and \mbox{IRAM$\,$30$\,$m}, both of which are used here as single-dish telescopes with a beam size of approximately $10\arcsec$ at 1.3\,mm,
are useful in supplementing the sky coverage of ALMA for the pulsar search (see Section~\ref{ssec:sensi_orb} for more details) and covering the position of PSR~J1745$-$2900.
The total length of the tracks on \mbox{Sgr$\,$A*} of these telescopes varied from 5 to 10\,hours on different nights, and are summarized in Table~\ref{tab:observations_summary}.


\begin{table*}
\centering
\caption{Details of observations and data analyzed in this paper. 
The first seven columns indicate the EHT station, the real (or equivalent for the phased ALMA array) single dish diameter ($D_{\rm s}$), the effective beam FWHM, the central observing frequency ($\nu_{\rm obs}$), the total effective bandwidth ($\Delta\nu$),
the number of frequency channels ($n_{\rm ch}$), and initial sample time ($t_{\rm samp}$), in the resulting PSRFITS files$^{\ddagger}$. The remaining columns show, for each epoch, the total time span of the dataset constructed for the pulsar search (i.e., the time between the start of the first scan and the end of the last scan) and the net integration time on \mbox{Sgr$\,$A*}. The difference between the total time span and the net integration time is due to the interleaved observations of calibrators, slew time of the telescope, time for pointing and focus adjustments and the flagged data. The data with the total time span are the ones searched, but only the net integration time on \mbox{Sgr$\,$A*} is considered for the search sensitivity analysis. 
}
\label{tab:observations_summary}
\begin{tabular}{lcccccccccccc}
\hline
\hline
\multicolumn{1}{c}{Station} & \multicolumn{1}{c}{$D_{\rm S}$} & \multicolumn{1}{c}{FWHM} & \multicolumn{1}{c}{$\nu_{\rm obs}$}  & \multicolumn{1}{c}{$\Delta\nu$}  & $n_{\rm ch}^{\ddagger}$ & $t_{\rm samp}^{\ddagger}$ & \multicolumn{3}{c}{Total time span (hr)} & \multicolumn{3}{c}{Net time on Sgr$\,$A* (hr)}\\
\cline{8-10} \cline{11-13}
 & \multicolumn{1}{c}{(m)} & \multicolumn{1}{c}{(\arcsec)} & (GHz) & (GHz) &  & $(\mu{\rm s})$ & Apr.~6 & Apr.~7 & Apr.~11 & Apr.~6 & Apr.~7 & Apr.~11 \\
\hline 
ALMA       & 74                         & 1$-$2 & 228.1 & 3.75$^*$ & 64 & 8 & 4.6 & 10.2 & 3.6 & 2.1 & 4.6 & 1.9\\
LMT        & 50$\,$(32.5$^{\dagger}$) & 10 & 228.1 & 4.00 & 32 & 8 & 5.7 &  6.4 & 3.6 & 2.4 & 3.0 & 1.7\\
\mbox{IRAM$\,$30$\,$m} & 30             & 10.8 & 228.1 & 4.00 & 32 & 8 & -   & 3.2  & -   & -   & 1.1 & -  \\
\hline 
\end{tabular}
\\ $^{\dagger}$\small{Though the full geometric diameter of LMT is 50$\,$m, during the EHT 2017 campaign only 32.5$\,$m were illuminated.}
\\ $^{\ddagger}$\small{During data preparation the number of frequency channels and sample time is reduced to varying degrees (see Section~\ref{ssec:data_convprep}).}
\\ $^*$\small{Due to slightly-overlapping channels in the digital filterbank, the effective bandwidth is 2$\times$1.875$\,{\rm GHz}$ \citep[see][]{goddi19}.}
\end{table*}

\subsection{Data conversion and preparation} \label{ssec:data_convprep}
The baseband voltage data were captured by \mbox{Mark$\,$6} recorders\footnote{\url{https://www.haystack.mit.edu/mark-6-vlbi-data-system/}} at the telescopes, and the disk packs shipped to the correlator at the Max Planck Institute for Radio Astronomy in Bonn, Germany, for post-processing. The voltage data selected for the pulsar search were reduced into multi-channel intensity time series (commonly refered to as a {\em filterbank}) written in search-mode PSRFITS  format, a standard FITS specification for pulsar data \citep{hotan04_psrfits}\footnote{\url{https://www.atnf.csiro.au/research/pulsar/psrfits_definition/PsrfitsDocumentation.html}}. This used the software tool \texttt{MPIvdif2psrfits}\footnote{\url{https://github.com/xuanyuanstar/MPIvdif2psrfits}}, an upgraded version of the original toolkit developed under the {\em ALMA Pulsar Mode Project}\footnote{\url{http://hosting.astro.cornell.edu/research/almapsr/}} \citep{lyw+19} that incorporates parallel processing capability using \texttt{OpenMPI}\footnote{\url{https://www.open-mpi.org/}}. The resulting properties of the PSRFITS products are presented in Table~\ref{tab:observations_summary}.




After conversion to PSRFITS the data consisted of a number of short scans 
for each of the two frequency sub-bands (high and low),
corresponding to  observations of \mbox{Sgr$\,$A*}, a calibrator, or another science target. From the scans on \mbox{Sgr$\,$A*}, those showing potential issues were flagged to be excluded from the analysis. The main reasons to flag scans were large variations or jumps of the mean power level, and, specifically for the case of ALMA, those scans not showing the expected array phasing noise features (see Section~\ref{sec:ALMAreduc}). 



Once the selection of \mbox{Sgr$\,$A*} scans to be analyzed were identified for ALMA, LMT and IRAM~30\,m, the data were further prepared for pulsar searching. The steps for each telescope are presented in the following subsections. 


\subsubsection{Atacama Large Millimeter/submillimeter Array}\label{sec:ALMAreduc}

When ALMA is used as a phased array the scans are partitioned into subscans with updates of the phase corrections every 18.192$\,$s to enable the coherent summation of signals from individual antennas \citep{goddi19}. This phasing cycle introduced a periodic feature in the time series, consisting in a decrease of the power level 
with a duration slightly below two seconds, followed by a large negative peak (we hereafter refer to this feature as the ``broad'' feature). Additionally, a forest of negative narrow spikes occur a few seconds after the broad feature \citep[see Appendix in][for details]{lde+21}. The first step in the preparation of the ALMA data was therefore to try to remove those noise features, which could introduce undesired power in the Fourier domain and decrease the sensitivity to pulsars in the search.
To locate and clean the features, we first produced a time series collapsing all the frequency channels of the PSRFITS file with the routine \texttt{prepdata} of the software package {\sc presto} \citep{ransom2011_presto_ascl}\footnote{\url{https://github.com/scottransom/presto}}, downsampling the time resolution to 32$\,\mu$s, and visually inspected each scan, marking the center of the first appearance of the broad feature. Then, two-second wide windows around the first broad feature position, and every 18.192$\,$s thereafter, were replaced with the mean level of the time series calculated outside the marked windows. 
Next, we use a moving window of width 10$\,$s that clips negative narrow spikes below a threshold of 5 times the standard deviation of the samples within the window.
The flagged data due to the phasing features correspond to $\approx$11\% of the total. 



Before combining the two frequency subbands (low and high) to obtain a full-bandwidth dataset, we remove slow variations of the mean power level in the time series with a ``running-fit'' filter that fits and subtracts a first order polynomial in a running window of 10$\,$s.  This step reduces the excess of power in the low-frequency end of the Fourier-domain data (also known as `red noise'), which tend to be large in mm- observations due to atmospheric opacity variations during observations.
We also normalize the time series by its standard deviation to combine the low and high subbands with a similar root mean square (RMS) level. At this stage, the cleaned time series of each individual scan and sub-band is saved as a {\sc sigproc}\footnote{\url{https://sigproc.sourceforge.net}}-format filterbank file \citep{sigproc_ascl} with one single frequency channel of 2 GHz and a time sampling of 32$\,\mu$s, except for April 7 where we downsample to 64$\,\mu$s to reduce the data size and speed up the processing of this particularly-long epoch. The production of the {\sc sigproc} filterbank files is made with custom Python code supported by tools from {\sc presto} and {\sc sigpyproc}\footnote{\url{https://github.com/ewanbarr/sigpyproc}}.

Then, for each sub-band, we concatenate all the scans for each observing track into a single, continuous {\sc sigproc} filterbank file,
padding the gaps in between scans with the mean value.
This concatenation is necessary to maintain the coherence of the data and maximize the sensitivity to periodic signals like those expected from pulsars.
Before the next step, we ensure that the start time and length are equal between the concatenated files in the low- and high-band datasets, adjusting the length if necessary. Finally, we splice together the two frequency sub-bands into one file with the \texttt{splice} routine of {\sc sigproc}. The result is a two-channel filterbank file containing all the scans of the observing track with the full bandwidth.
This process is repeated for each night, and the final filterbank files are the ones passed to the searching pipelines (see Section~\ref{sec:searches}). 

In a separate step, we Fourier transformed and analyzed with {\sc presto}'s \texttt{accelsearch} routine 
three scans on calibrators (one for each observing night) to obtain a list of locally-generated periodic signals through an analysis of the corresponding Fourier spectrum. A list of periodicities --- most likely produced in the receiving chain --- is created, and later used to flag them when applying the searching pipeline to the observations on \mbox{Sgr$\,$A*} by zero-weighting the corresponding Fourier bins. 


\subsubsection{IRAM 30m Telescope}\label{sec:30mreduc}

The IRAM 30m telescope data were mainly affected by two features: a strong ripple in the time series with a frequency of about 210.7$\,$Hz, 
and regular power drop-offs, part of them synchronized with the 210.7$\,$Hz cycle (see Figures~\ref{fig:ts_examples} and \ref{fig:ts_examples_ZOOM} in the Appendix). The origin of these noise features is under investigation, suspected to be related to oscillations in the bias circuits of the first mixer of the receiver. To reduce the impact in the subsequent analysis, we first zero-weight a few Fourier bins centered on the 210.7$\,$Hz signal, including 3 harmonics. We also zero-weight a few bins around 1$\,$Hz, to remove another strong periodic signal related to the cryogenic pump cycle. To reduce the number of power drop-offs, we visually inspect the resulting time series, and manually mark a threshold below which all samples are substituted with the median of the remaining data. The percentage of data flagged related to the power drop-offs is on average 2.8\%. 
The following steps are analogous as for the ALMA data: a running-fit filter with a window of 10$\,$s to subtract slow mean level variations, a normalization of the data by their standard deviation, and a concatenation of scans plus splicing into a single filterbank file. Five scans showing larger instabilities in the form of ripples or considerably more signal drop-offs than on average, amounting to 0.72$\,$hr in total, were flagged and excluded from the analysis.


\subsubsection{Large Millimeter Telescope}\label{sec:LMTreduc}

The data from the LMT show strong periodicities associated with the local receiver. The most prominent signals have a frequency of 1.2$\,$Hz, probably related to the cryogenics 
refrigeration cycle, and around 185$\,$Hz, which has an unknown origin. 
The data also show some power drop-offs, but less frequently than the ALMA and \mbox{IRAM$\,$30$\,$m} data. A variation of the mean power level is usually present, related to opacity variations during the observations (see Figures~\ref{fig:ts_examples} and \ref{fig:ts_examples_ZOOM} in the Appendix). To reduce these undesired signals, we zero-weight the strongest peaks related to the 1.2 and 185$\,$Hz signals in the Fourier domain, and after an inverse Fourier transform, use a moving window of 10$\,$s that clips the power-drops substituting them with the median value inside the window. The mean power level variations are minimized using the running-fit filter with a window of 10$\,$s as done for the ALMA and \mbox{IRAM$\,$30$\,$m} data. The remaining steps --- normalization, concatenation and splicing --- are analogous to the ALMA and \mbox{IRAM$\,$30$\,$m} datasets. For the LMT, 16 scans in total (amounting to 2.32$\,$hr) were flagged because they still showed clear artifacts after the cleaning, like jumps in the mean power level and regular trains of strong, wide-pulse-like features.



We remark that although the undesired signals are highly reduced in the data from the three stations by our cleaning algorithms, they are not fully removed. We show examples of the time series before and after the cleaning procedures in Appendix~\ref{sec:appA}.

\subsection{Pulsar searching}\label{sec:searches}

Searching for pulsars via their inherent periodicity is done with two independent pipelines: a Fourier-domain search \citep{ranein2002} and a search using the Fast-Folding-Algorithm \citep[FFA,][]{staelin1969fast}. Fourier-domain methods are widely used in pulsar searches and have proven successful in discovering pulsars with a variety of spin parameters \citep{lyne+03}. The FFA works particularly well at detecting long-period pulsars that often show narrow pulse profiles \citep{2017MNRAS.468.1994C,2020MNRAS.497.4654M}. In addition, we searched the data for single-pulse burst-like transient emission\footnote{cf. \citet{musetal22} that carried out a search for transients with the ALMA data but at longer ($\sim$seconds to minuntes) time scales and therefore not sensitive to the same emission sources.} as seen in pulsar giant pulses, Rotating Radio Transients (RRATs) and Fast Radio Bursts (FRBs) \citep[see~e.g.][]{keane2011}. In the following subsections, we detail the search algorithms used. 
The parameters utilized for each pipeline are summarized in Tables~\ref{tab:searchparams_and_candidates} and \ref{tab:spsearch_params_results}. 

Pulsar surveys at low radio frequencies typically search over a range of dispersion measures (DM) --- the integrated column density of free electrons along the line-of-sight. However, there is no dedispersion in our pipelines because the effect of even large DMs
is negligible at the EHT observing frequencies and for the target time resolution of 32$\,\mu$s. 
Interstellar dispersion delays the pulses' arrival times scaling with $\rm{DM}\,\nu^{-2}$ \citep[see e.g.,][]{lorkra04}, but a DM=10000 pc$\,$cm$^{-3}$ would produce a pulse smearing of only $\Delta t_{\rm DM}\simeq28\,\mu$s across our $\sim$4-GHz band. The highest DM known to date is for the magnetar PSR~J1745$-$2900, located close to \mbox{Sgr$\,$A*}, at DM=1778 pc$\,$cm$^{-3}$ \citep{eat13}. The negligible dispersion has some disadvantages (see Sections~\ref{sec:sp_analysis}, \ref{sec:disc_cleaning}), but reduces considerably the computing costs, which can be concentrated on the acceleration search. 





\subsubsection{Periodicity search in the Fourier domain}\label{sec:FFTsearch}

The Fourier-domain pipeline predominantly searches for pulsars by detecting and analysing peaks in the Fourier spectrum of the time series.
We use the software {\sc presto}, which includes the capability to search for accelerated pulsars, like those orbiting \mbox{Sgr$\,$A*} or other companions. The acceleration parameter space is searched using a template-matching algorithm by recovering the power spread over contiguous Fourier bins \citep{ranein2002}. This power spread is the result of the Doppler effect when a pulsar changes its radial velocity during an observation. 

A limitation of the template-matching algorithm is that it loses efficiency for observations covering a substantial fraction of the binary period ($P_{\rm b}$). This limiting fraction is about $0.1P_{\rm b}$ (given a companion mass of 1.4 solar mass), i.e., the algorithm works well when the total observing time does not exceed about $10\%$ of the binary period \citep[e.g.,][]{ransom2001,ncb+15}. 
To improve the sensitivity for longer observing spans, one can add a search in the line-of-sight (LoS) acceleration derivative (commonly referred to as ``jerk'') parameter space \citep{and18}, although this increases greatly the computational costs\footnote{The number of computations scales roughly proportionally to the square of the total observing time for searching acceleration, and with the cube when adding a jerk search.}.
The LoS acceleration ($a_{\rm l}$) and its derivative ($j_{\rm l}$) of a pulsar in orbit can be calculated as \citep[e.g.,][]{blw13}: 
\begin{equation}
a_{\rm l}=-\left(\frac{2\pi}{P_{\rm b}}\right)^2\frac{a_{\rm p}\sin i}{(1-e^2)^2}\sin(A_{\rm T}+\omega)(1+e\cos A_{\rm T})^2,
\end{equation}
and
\begin{eqnarray}
j_{\rm l}&=&-\left(\frac{2\pi}{P_{\rm b}}\right)^3\frac{a_{\rm p}\sin i}{(1-e^2)^{7/2}}(1+e\cos A_{\rm T})^3\cdot[\cos(A_{\rm T}+\omega) \nonumber\\
    &&+e\cos\omega-3e\sin(A_{\rm T}+\omega)\sin A_{\rm T}],
\end{eqnarray}
where $a_{\rm p}$, $i$, $e$, $\omega$ and $A_{\rm T}$ are the semi-major axis, inclination angle, eccentricity, longitude of periastron and true anomaly of the orbit, respectively.


Our {\sc presto}-based pipeline is based on the one used by \citet{tor21}. It first produces a time series from the filterbank file and then uses a Fast Fourier Transform (FFT) to transform the data to the frequency domain. Next, it searches for periodic signals as discussed above. To increase the sensitivity to long-period pulsars, the \texttt{rednoise} filtering routine of {\sc presto} is applied to further decrease the red noise. Up to ${\rm N_{h}}=32$ harmonics of each detected periodicity are summed. 

To be sensitive to a wide range of different pulsar systems, two passes on each observation are made. In the first pass, highly accelerated pulsars are searched, and a second pass searches for accelerated pulsars including a search in the jerk parameter space. In both cases searches for isolated pulsars (i.e., with no acceleration) are included. The control of the ranges for surveying acceleration and jerk are made through the parameters \texttt{zmax}=max($|z|$) and \texttt{wmax}=max($|w|$) of the routine \texttt{accelsearch}, which in turn represent the maximum shift in the Fourier frequency bin ($z$) and its derivative ($w$) within an entire observation of length $T_{\rm obs}$ that the pipeline will search. Following \citet{and18}, these two terms can be written as:
\begin{equation} \label{eq:zmax}
    z=\frac{a_{\rm l}hfT_{\rm obs}^2}{c}, 
\end{equation}
and
\begin{equation} \label{eq:wmax}
    w=\frac{j_{\rm l}hfT_{\rm obs}^3}{c}, 
\end{equation}
where $f$, $c$, $h$ represent the pulsar rotational frequency, speed of light, and the index of the Fourier harmonic. Here, we use \texttt{zmax}=1200 and \texttt{wmax}=0 for the pass aiming for highly accelerated pulsars,
and \texttt{zmax}=300 and \texttt{wmax}=900 for the pass including a jerk search. For isolated pulsars both parameters are set to zero. 

The exact relationship between sensitivity, pulsar spin period, acceleration (or orbital period), jerk, pipeline parameters, and observing time is complex \citep[see e.g.,][]{lde+21, etd+21}. Nonetheless, with the parameters \texttt{zmax} and \texttt{wmax} chosen here, we can correct for the LoS acceleration of virtually any pulsar orbiting \mbox{Sgr$\,$A*} down to orbital periods $P_{\rm b}\geq2.5\,$yr (assuming circular orbits). Pulsars in orbits with shorter orbital periods can still be detected, depending on their strength and spin period. A more detailed discussion on the coverage of pulsar orbits by our search is presented in Section~\ref{ssec:sensi_orb}.

After all the pulsar candidates found by \texttt{accelsearch} are saved, a sifting step removes duplicated and harmonically related ones. This step admits pulsar candidates with $\sigma_{\rm sift}\geq2.0$ as calculated by the \texttt{sifting.py} code\footnote{In general, {\sc presto}'s $\sigma$ values provides a way to estimate if a signal is due to noise or not. While the powers are $\chi^2$ distributed, {\sc presto} makes a conversion to equivalent gaussian significance. The parameter $\sigma_{\rm sift}$ therefore indicates the probability that a signal has of not being produced just by the noise. The code takes into account the number of trials in the search, normalizing to a ``single-trial'' probability.}. Such a $\sigma_{\rm sift}$ threshold allows for the detection of weak pulsars while maintaining a manageable number of candidates \citep[see][and Section~\ref{sec:pipeverification}]{tor21}. A final step uses \texttt{prepfold} to fold\footnote{Folding is a common technique used in high-time-resolution observations of pulsars in which a long observation is split in blocks of length equal to the spin period of a target pulsar. Then, the blocks are summed or averaged together. The result, in case of a detection, is an averaged profile of the pulsar emission during the observation, together with a substantial improvement in sensitivity due to increase of signal-to-noise ratio by the addition of the pulsar signal in each block.} the data with the information of each candidate, producing plots that can be inspected to decide if any of the candidates corresponds to a real pulsar. 

The folding step produces four plots from each candidate. Two of these plots correspond to the raw data with and without an optimization of candidate parameters from \texttt{prepfold}. The other two plots arise from a filtered version of the raw data using the \texttt{rednoise} filter from {\sc presto}. The \texttt{rednoise} filter is very effective in  reducing some local,  interfering periodic signals in some  datasets, and can in some cases be key to enabling a detection \citep[for an example, see Figure~2 in][]{tor21}. This multiple candidate plot production, in exchange of increasing the number of candidate plots to inspect, decreases the risk of missing weak pulsars by an insufficient cleaning or by interference from the locally-generated periodic signals.

\subsubsection{Periodicity search with Fast Folding Algorithm}\label{sec:FFAsearch}

Unlike the Fourier-domain pipeline, the FFA searches for pulsars by folding the time series at different trial periods, forming a sequence of profiles, and testing the significance of each profile using boxcar matched filters. The pulse width and arbitrary total pulse power produced by the filters are used to calculate the signal-to-noise ratio (S/N$_{\rm FFA}$) of the pulse profiles
\citep[see e.g.,][for details]{2017MNRAS.468.1994C}. We used the software \textsc{riptide}\footnote{\url{https://github.com/v-morello/riptide}} \citep{2020MNRAS.497.4654M} as the basis for the FFA search pipeline. 
We include an acceleration search by resampling the time series using {\sc sigpyproc} at a series of acceleration trials using a custom pipeline (Wongphechauxsorn et al., \emph{in prep.}). 
This implementation of the FFA pipeline assumes a constant acceleration during an observation. 

The \textsc{riptide} software uses a matched filter to detect the pulse in the folded data; making the number of folding profile bins ($\rm N_{bin}$) an important parameter. If the duty cycle is less than one bin, the sensitivity will be reduced. \citet{2020MNRAS.497.4654M} demonstrated that the number of trials is proportional to $\rm N_{bin}^3$. Folding with many profile bins consequently results in more trials and longer processing times. Furthermore, the acceleration step between trials is proportional to $\rm N_{bin}$,
thus the total number of trials is proportional to $\rm N_{bin}^4$.

Our FFA pipeline uses the cleaned time series generated following the description in Section~\ref{ssec:data_convprep}. We applied the FFA acceleration search to the time series for periods from $P_{\rm min}=1\,$s to $P_{\rm max}=1025\,$s, using $\rm N_{bin}=128$. As a result, the pipeline provides full sensitivity to duty cycles down to about 1\%\footnote{The duty cycle was calculated using the period pulse width from PSRCAT \citep{mhth05}. N.b., less than 7\% of known pulsars have a lower duty cycle.}). We used an acceleration range of $a_{\rm l}=\pm$50$\,$m s$^{-2}$, i.e., approximately equivalent to \texttt{zmax}=1400 for detecting a 1-s pulsar in the Fourier domain acceleration search pipeline with $\rm{N_h} = 32$ and $T_{\rm obs}$ of 4.5\,hr. The FFA can therefore probe the same type of binary orbit as the Fourier-domain acceleration search for long-period pulsars.
Furthermore, the FFA pipeline is sensitive to an acceleration derivative up to approximately $2.6\times10^{-5}\,$m s$^{-3}$ without needing any jerk search. The range of acceleration  derivative covered is equivalent approximately to a \texttt{wmax}=10 for a 1-s pulsar in the Fourier domain acceleration search pipeline with the same parameters ($\rm{N_h} = 32$ and $T_{\rm obs}=4.5\,$hr). 

In addition, to search for very narrow duty cycles, the FFA was repeated using  $\rm N_{bin}=512$ with no acceleration, meaning that the pipeline is sensitive to pulsars with a duty cycle larger than 0.19\% (only $\approx0.2\%$ of currently-known pulsars have a lower duty cycle). All candidates detected with S/N $ \geq $ 6 generated by either the accelerated or non-accelerated pipeline were folded to be inspected visually.

\subsubsection{Single pulse search}\label{sec:sp_analysis}
While a pulsar signal is often characterized as a series of stable pulsations, individual single pulses --- which can be orders of magnitude brighter than the averaged pulse emission --- are observed in some cases \citep[][]{mc03} and are detectable with alternative search techniques \citep[][]{cm03}. For this task we employed \texttt{single\_pulse\_search.py} in {\sc presto}. This routine identifies pulses by matched-filtering the time series with boxcar filters of different pulse widths up to a maximum of 300 samples; corresponding to 9.6$\,$ms and 19.2$\,$ms for our datasets re-sampled at 32$\,\mu$s and 64$\,\mu$s, respectively. To avoid excessive automatic time series flagging, we adjusted the  \texttt{single\_pulse\_search.py} internal parameters \texttt{lo\_std} and \texttt{hi\_std} to $0.88$ and $3.68$ respectively. 
For the EHT data at $1.3\,{\rm mm}$ pulse dispersion is negligible, so it cannot be used as the marker of a genuine astrophysical pulse. Instead the simultaneous nature of observations conducted at multiple sites is utilized to match coincident pulses from each telescope in the common rest frame of the solar system barycentre.     

In practice \texttt{single\_pulse\_search.py} gives information on the single pulse significance, $\sigma_{\rm SP}$, its arrival time along with the corresponding number of samples relative to the beginning of the observation and the pulse width (determined by the optimum boxcar filter size). 
By transforming all the single pulse arrival times from each site to their equivalent barycentric Modified Julian Date (MJD) $-$ determined from {\sc presto} using the NASA Jet Propulsion Laboratory planetary ephemeris DE405\footnote{\url{https://ssd.jpl.nasa.gov/planets/eph_export.html}} in {\sc tempo} \citep{tempo_ascl}\footnote{\url{https://tempo.sourceforge.net}} $-$ we search for pulses that are coincident to an accuracy of one time sample (32 or 64$\,\mu$s depending on dataset).

In addition, any pulse with $\sigma_{\rm SP}>12$ is visually inspected (regardless of coincidence)
to check for other astrophysical markers such as scatter broadening of the profile or `multi-component profiles' as seen in many pulsars. This also accounts for different \mbox{Sgr$\,$A*} visibility windows at each station, their various individual single-pulse sensitivities, and the different spatial coverages due to different beam sizes (see Tables~\ref{tab:observations_summary}~and~\ref{tab:spsearch_params_results}).

\subsection{Verification of periodicity search pipelines with synthetic signal injections}\label{sec:pipeverification}

Similarly to the method in \cite{tor21}, before searching the data a number of pre-processing tests were carried out to verify and tune the pipelines. This was done by injecting synthetic pulsars into the real data and checking their correct detection. The signals were produced with a custom-version of {\sc sigproc}'s \texttt{fake} and injected to the filterbank files. The synthetic pulsars included slow-spinning pulsars ($P\geq2\,$s), canonical pulsars ($P\sim500\,$ms), and fast-spinning millisecond pulsars (MSPs; $1\,\rm{ms}<P\leq30\,$ms). Within each category, different intensities and duty cycles were used, and both isolated  and binary pulsars in a range of configurations were injected. The simulated companions included neutron stars, stellar-mass black holes and a supermassive black hole with the mass of Sgr$\,$A*.

The tests served two main purposes: first, to allow fine adjustments of the pipeline parameters so that injected signals were recovered and secondly, to verify that even extreme pulsar systems, like MSPs in tight orbits around massive companions, were detected when the characteristics are within the theoretical limitations of the searching algorithms (see~Section~\ref{ssec:sensi_orb}). Here we confirmed the need of a low value of {\sc presto}'s \texttt{sifting.py} sigma threshold, which is a significance threshold to accept or reject candidates after a sifting step (set to $\sigma_{\rm sift}>2.0$), together with a weak requirement on the minimum number of harmonics required per candidate, that is set to one \citep{tor21}.

\section{Results} \label{sec:results}

\subsection{Periodicity search}\label{sec:results_periodicity}

\subsubsection{Fourier-domain acceleration search}\label{sec:results_presto}


The analysis with the Fourier-domain pipeline of the three epochs of ALMA resulted in 3146 pulsar candidates. 
However, no new pulsars are discovered. We identify a large number of candidates with a round number in the spin period or frequency (e.g. $\nu_{\rm s} =$ 437.500000, 111.111111, 2875.000000$\,$Hz, etc.). Another remarkable characteristic of the detected signals is a very short period, with 
56.3\% of the candidates having $P<0.5\,$ms. Those characteristics, in particular the round numbers, suggest a
human-generated origin. Furthermore, we observe similar signals in the analysis of off-source scans (see Section~\ref{sec:ALMAreduc})
and we therefore relate these found periodicities to the local receiving chain and the properties of the Superconductor-Insulator-Superconductor (SIS) receivers used in the observations \citep[see also][]{tor21}. 





The single epoch with the \mbox{IRAM$\,$30$\,$m} telescope resulted in 99 pulsar candidates.
No signal resembling a real pulsar was found. The candidates are similar in properties to those detected for ALMA, with a significant number showing round values in frequency, and the majority with short periods, with $P\leq4\,$ms. The \mbox{IRAM$\,$30$\,$m} telescope used also a SIS receiver, and similarly to the case of ALMA, we relate the candidates to local oscillations inside the receiver or data transport chains. 


Searching the three epochs from the LMT yielded 984 candidates. 
One candidate stands out after the analysis, with a spin period very close to 32$\,$ms. After a careful examination we conclude that this candidate is related to locally-generated periodic signals or a digitization artifact, because the spin period is almost exactly a thousand times the sampling time of 32$\,\mu$s. Other reasons to classify this candidate as local are its high power, the fact that the signal suffers at least one significant jump in rotational phase during two of the three observed epochs, and because the same signal is not detected in the ALMA nor the \mbox{IRAM$\,$30$\,$m} datasets. Other strong periodicities at 100, about 180, and 200$\,$Hz are detected in the LMT data, likely related to the power supply. Similarly to the other stations, the data show a significant number of candidates found at round periodicities (both in period or frequency). We conclude that no new pulsars are discovered in the LMT data. 

\begin{figure*}
    \centering
    \subfloat{
         \centering
         \includegraphics[width=\textwidth]{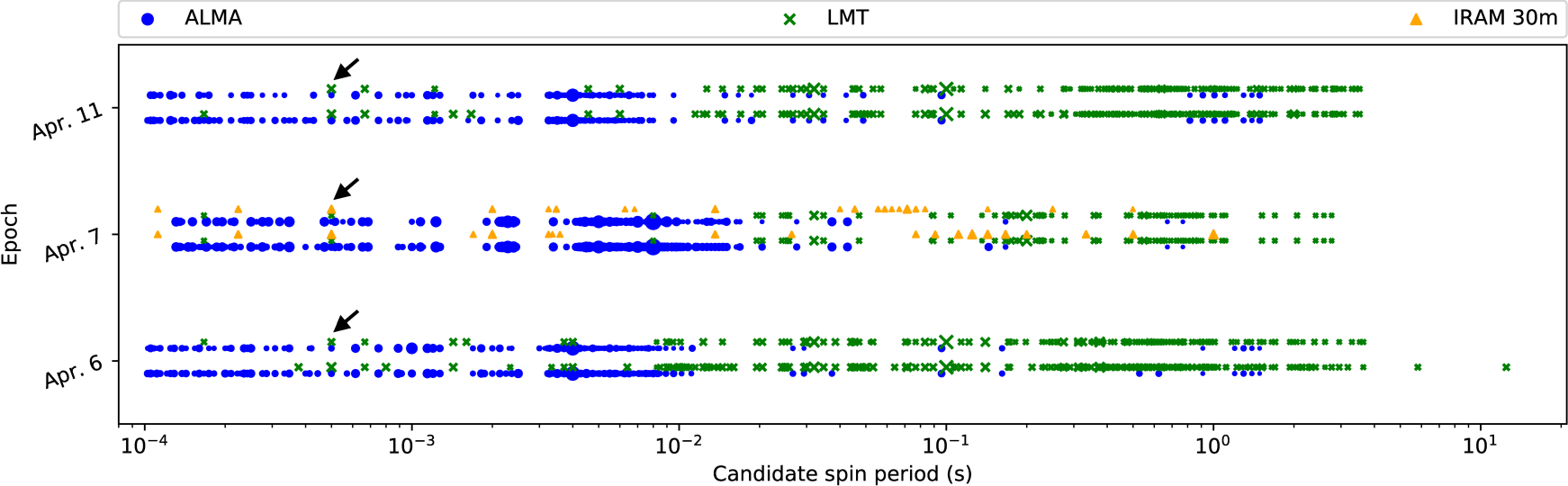}
         \vspace*{0.1cm}
    }\\
    \subfloat{
         \centering
         \includegraphics[width=\textwidth]{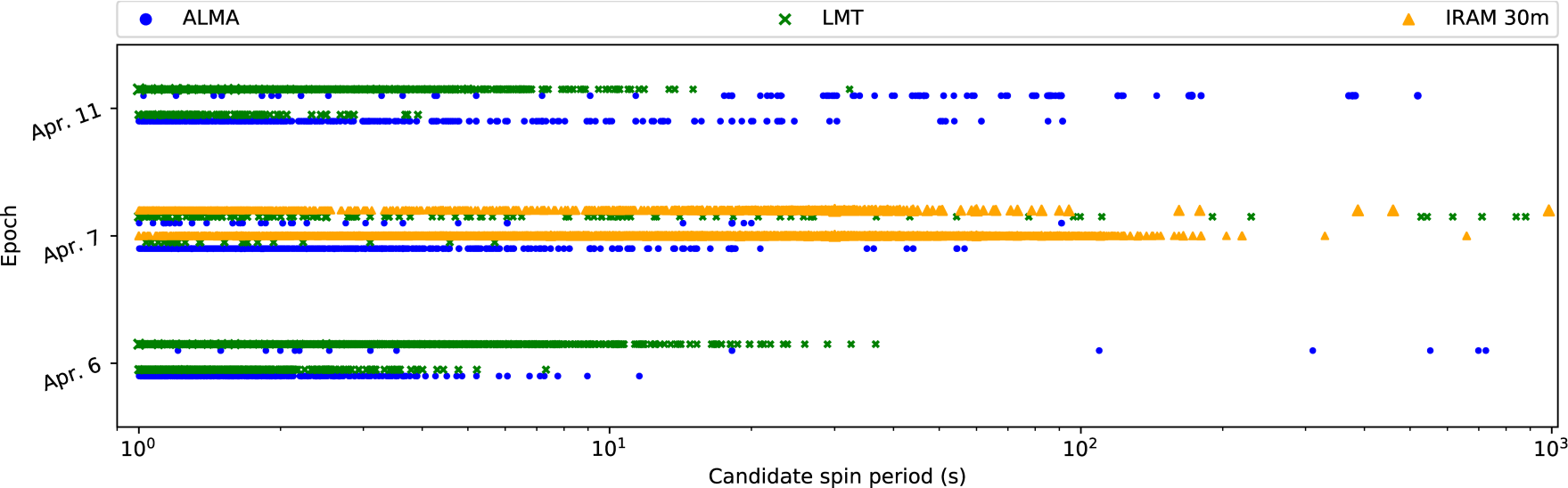}
    }
    \caption{Spin period of the candidates from the two periodicity-search pipelines: Fourier-domain analysis (top panel) and FFA (bottom panel). 
    Each epoch shows two lines of markers, corresponding to the two data processing passes with each pipeline (labeled as FFT$_1$, FFT$_2$, FFA$_1$ and FFA$_2$ in Table~\ref{tab:searchparams_and_candidates}). The blue circles, green crosses, and orange squares represent the candidates from ALMA, LMT and the \mbox{IRAM$\,$30$\,$m} telescope, respectively. The size of the marker is proportional to the $\sigma_{\rm sift}$, and S/N ratio values, for the Fourier-domain and FFA cases, respectively. None of the candidates is a convincing pulsar signal after the inspection of their candidate evaluation plots. The black arrows in the top panel indicate the $P=0.5\,$ms candidate, the only periodicity present in all datasets and related to a locally-generated signal at $\nu=2000\,$Hz (see Section~\ref{sec:res_comparison}). \label{fig:candidates}}
\end{figure*}

Table \ref{tab:searchparams_and_candidates} provides details of the number of candidates from each epoch, station, and searching pass, and we represent visually the spin period of the candidates in Figure~\ref{fig:candidates}.

\subsubsection{Fast-Folding-Algorithm with acceleration search}\label{sec:results_FFA}


\par The FFA pipeline applied to the ALMA data resulted in 3044 candidates.
The strongest candidate had a period around 18.2$\,$s, which is close to the feature found in the raw ALMA data due to the phasing cycle. The rest of the candidates have a period close to the harmonics of this signal. This candidate and its harmonics likely originates from residual contamination from the known 18.2-s period phasing features.

\par The FFA pipeline on the LMT data resulted in 300 candidates. 
Most of the candidates with $\rm S/N > 7$ 
showed repetitions of the pulse in the folded profile, indicating that the fundamental periods of these signals are below our period search range. The candidates also have periods with highly-round numbers, e.g., 1.0000\,s, 2.5000\,s, and 1.6000\,s. We conclude that these candidates have a local origin.
\par For the \mbox{IRAM$\,$30$\,$m} the FFA search with acceleration yielded over 16000 candidates. This number is much higher than the number of candidates detected for ALMA and the LMT, indicating that our S/N threshold is likely too low resulting in too many false candidates.
To limit the number of accepted candidates, the minimum S/N ratio was increased to 7, 
reducing the candidates to 4185 while still maintaining a low significance limit.


\par A strong 
candidate at 3.0$\,$s was found without acceleration, with detections of harmonics (i.e., period of 6.000$\,$s and even 150.0$\,$s). This candidate's period is close to an integer to the third decimal, suggesting that it is not astrophysical.
We confirmed that this candidate likely originates locally by a repeat of the search in topocentric frame, which yields the same candidate with an exact period of 3.000$\,$s.


We conclude that no pulsars were found by the FFA pipeline. Table~\ref{tab:searchparams_and_candidates} includes a summary of the candidates, and Figure~\ref{fig:candidates} a visual presentation of the candidates' spin periods.

\begin{table*}
	\centering
    \caption{Summary of the parameters of the two periodicity-search pipelines. Columns show the station, epoch, pipeline (with subscript indicating different passes with different parameters), parameters used in each case (see Sections~\ref{sec:FFTsearch} and \ref{sec:FFAsearch} for details), and number of pulsar candidates produced in each search pass. A ``$-$'' symbol indicates that for that pipeline the parameter is not applicable.}
	\label{tab:searchparams_and_candidates}
	\begin{tabular}{cccccccccccccc} 
		\hline
		Station   & Epoch   & Pipeline & \multicolumn{9}{c}{Parameters}                                                                                              & N$_{\rm cands}$\\
		\cline{4-12} 
		            &         &               & \texttt{zmax} & \texttt{wmax} & ${\rm N_{h}}$ & $\sigma_{\rm sift}$ & $P_{\rm min}$ & $P_{\rm max}$ & $a_{\rm l}$             & $\rm N_{bin}$ & $\rm S/N_{FFA}$ &      \\
		            &         &               &      &      &                  &          & (s)           & (s)           & ($\rm m/s^2$) &             &                   &               \\ 
		\hline
		ALMA        & 2017 April 6 & FFT$_1$  & 1200 & 0   & 32                & $>$2.0      & $-$           & $-$           & $-$           & $-$         & $-$               &  539      \\
		            &              & FFT$_2$  & 300  & 900 & 32                & $>$2.0      & $-$           & $-$           & $-$           & $-$         & $-$               &  1043     \\
		            &              & FFA$_1$  & $-$  & $-$ & $-$               & $-$      & 1             & 1025          & 0             & 512        & $\geq$ 6.0               &  16      \\
		            &              & FFA$_2$  & $-$  & $-$ & $-$               & $-$      & 1             & 1025          & $\pm50$       & 128        & $\geq$ 6.0               &  556       \\   
		            
		            & 2017 April 7 & FFT$_1$  & 1200 & 0   & 32                & $>$2.0      & $-$           & $-$           & $-$           & $-$         & $-$               & 175\\
		            &              & FFT$_2$  & 300  & 900 & 32                & $>$2.0      & $-$           & $-$           & $-$           & $-$         & $-$               & 341\\
		            &              & FFA$_1$  & $-$  & $-$ & $-$               & $-$      & 1             & 1025          & 0             & 512        & $\geq$ 6.0               & 37\\
		            &              & FFA$_2$  & $-$  & $-$ & $-$               & $-$      & 1             & 1025          & $\pm50$       & 128        & $\geq$ 6.0               & 1725          \\
		            
		            & 2017 April 11 & FFT$_1$  & 1200 & 0   & 32                & $>$2.0      & $-$           & $-$           & $-$           & $-$         & $-$               & 315\\
		            &               & FFT$_2$  & 300  & 900 & 32                & $>$2.0      & $-$           & $-$           & $-$           & $-$         & $-$               & 733\\
		            &               & FFA$_1$  & $-$  & $-$ & $-$               & $-$      & 1             & 1025          & 0             & 512        & $\geq$ 6.0               & 91          \\
		            &               & FFA$_2$  & $-$  & $-$ & $-$               & $-$      & 1             & 1025          & $\pm50$       & 128        & $\geq$ 6.0               & 619          \\   
		            
		LMT         \rule{0pt}{2.5ex} & 2017 April 6 & FFT$_1$ & 1200 & 0   & 32                & $>$2.0      & $-$           & $-$           & $-$           & $-$         & $-$     & 162 \\
		            &               & FFT$_2$  & 300  & 900 & 32                & $>$2.0      & $-$           & $-$           & $-$           & $-$         & $-$               & 262 \\
		            &               & FFA$_1$  & $-$  & $-$ & $-$               & $-$      & 1             & 1025          & 0             & 512        & $\geq$ 6.0               & 43           \\
		            &               & FFA$_2$  & $-$  & $-$ & $-$               & $-$      & 1             & 1025          & $\pm50$       & 128        & $\geq$ 6.0               & 512          \\   
		            
		            & 2017 April 7  & FFT$_1$  & 1200 & 0   & 32                & $>$2.0      & $-$           & $-$           & $-$           & $-$         & $-$               & 67 \\
		            &              & FFT$_2$  & 300  & 900 & 32                & $>$2.0      & $-$           & $-$           & $-$           & $-$         & $-$               & 92 \\
		            &              & FFA$_1$  & $-$  & $-$ & $-$               & $-$      & 1             & 1025          & 0             & 512        & $\geq$ 6.0               & 10          \\
		            &              & FFA$_2$  & $-$  & $-$ & $-$               & $-$      & 1             & 1025          & $\pm50$       & 128        & $\geq$ 6.0               & 26          \\   
		            
		            & 2017 April 11 & FFT$_1$ & 1200 & 0   & 32                & $>$2.0      & $-$           & $-$           & $-$           & $-$         & $-$               & 139 \\
		            &              & FFT$_2$ & 300  & 900 & 32                & $>$2.0      & $-$           & $-$           & $-$           & $-$         & $-$               & 262 \\
		            &              & FFA$_1$  & $-$  & $-$ & $-$               & $-$      & 1             & 1025          & 0             & 512        & $\geq$ 6.0               &    50       \\
		            &              & FFA$_2$  & $-$  & $-$ & $-$               & $-$      & 1             & 1025          & $\pm50$       & 128        & $\geq$ 6.0               & 145          \\   
		            
		\mbox{IRAM$\,$30$\,$m} \rule{0pt}{2.5ex} & 2017 April 11 & FFT$_1$ & 1200 & 0   & 32                & $>$2.0      & $-$           & $-$           & $-$           & $-$         & $-$ & 33 \\
		            &              & FFT$_2$ & 300  & 900 & 32                & $>$2.0      & $-$           & $-$           & $-$           & $-$         & $-$               &  66\\
		            &              & FFA$_1$  & $-$  & $-$ & $-$               & $-$      & 1             & 1025          & 0             & 512        & $\geq$ 6.0               & 454           \\
		            &              & FFA$_2$  & $-$  & $-$ & $-$               & $-$      & 1             & 1025          & $\pm50$       & 128        & $\geq$ 7.0               & 4179         \\   
		\hline
	\end{tabular}
\end{table*}

\subsubsection{Comparison of candidates among stations}\label{sec:res_comparison}



Apart from an individual review of all candidates, we cross-checked repetitions of periodicities between epochs and stations. For the Fourier-domain pipeline, we use a similar method as for the single pulse coincidence matching (see Section~\ref{sec:sp_analysis}) based on \texttt{numpy\_intersect\_1D}. To match candidates with periodicities that may not coincide exactly (this can occur for example by a slightly different result of optimization of parameters from \texttt{prepfold}), we cross-check the periodicities with precision down to 100$\,\mu$s. From the 442 reviewed coincidences, none resembles a real pulsar signal;  the results are again dominated  by local round-number periodicities. 

In the few cases where repetitions of periodicities with no round values are found, the characteristics of the candidate (e.g., significance or profile shape) made us classify them as false, produced again likely by local signals or the noise. The most repeated candidate is this step was the $P=0.5\,$ms, identifiable in Figure~\ref{fig:candidates}. This P=0.5$\,$ms candidate is clearly detected in the three stations at all epochs. It is however related to a local signal at a frequency of $\nu_{\rm s}$=2000$\,$Hz. The fact that it exists in all of the datasets reinforces a relationship with the EHT-specific backend (e.g., its clock signal generator, digitizers and/or \mbox{Mark$\,$6} recorders), that were common hardware at the three telescopes. For the FFA pipeline the periodicities were compared to within 1\% of precision, yielding no coincidences between the candidates.

\subsection{Single pulse search}\label{sec:SPresults}

The single-pulse pipeline resulted in a total of 6893 candidate pulses; the majority of which we determine to be erroneous events. Table~\ref{tab:spsearch_params_results} shows the number of pulses detected per station  epoch, $\sigma_{\rm SP}$ thresholds and representative flux density limits. 
The large discrepancy in pulse numbers
corresponds to a combination of the 
non-Gaussian noise properties of each dataset and the different observing lengths. 
The ALMA data on April 6 and 11 yield the lowest number of single-pulse candidates, however, on April 7 where there are more instrumental instabilities and residual artifacts in the time series (see Section~\ref{ssec:data_convprep} and Appendix~\ref{sec:appA}), combined with the longer integration length, a significantly higher number of candidates is observed.
In general the LMT and the IRAM~30\,m datasets also produced larger numbers of candidates for the same reasons and despite the higher $\sigma_{\rm SP}$ thresholds used. 




\begin{table}
\centering
\caption{Parameters and results of the search for single pulses and transients. Columns show the dataset analysed (date and station), the signal-to-noise threshold used for single pulse events ($\sigma^{\prime}_{\rm SP}$),  the number of single pulses detected ($\rm N_{\rm cand}^{\rm SP}$), and flux density limits ($S_{\rm min}$) for single pulses of representative widths $1, 10$ and $50\,{\rm ms}$.}
\label{tab:spsearch_params_results}
\begin{tabular}{lrrrrr}
\hline
\hline
\multicolumn{1}{c}{Dataset}    &    \multicolumn{1}{l}{$\sigma^{\prime}_{\rm SP}$}  &  \multicolumn{1}{c}{$\rm N_{\rm cand}^{\rm SP}$} & \multicolumn{1}{c}{$S^{\rm 1\,ms}_{\rm min}$} &  \multicolumn{1}{c}{$S^{\rm 10\,ms}_{\rm min}$}  & \multicolumn{1}{c}{$S^{\rm 50\,ms}_{\rm min}$} \\
& &  & \multicolumn{1}{c}{(Jy)} & \multicolumn{1}{c}{(Jy)} & \multicolumn{1}{c}{(Jy)} \\
\hline 
Apr.\,6,  ALMA        & 6     &     10 &  0.26 &  0.08  & 0.03  \\
Apr.\,6,  LMT         & 10    &    574 & 38.04 & 12.03  & 5.38  \\

Apr.\,7,\,ALMA        & 6     &   1231 &  0.25 &  0.08  & 0.04   \\
Apr.\,7,\,LMT         & 10    &   1125 & 27.51 &  8.70  & 3.89   \\
Apr.\,7,\,IRAM~30\,m  & 10    &   3822 &  6.41 &  2.03  & 0.91   \\

Apr.\,11,\,ALMA       & 6     &     68 &  0.41 & 0.13 & 0.06    \\
Apr.\,11,\,LMT        & 10    &     63 & 23.01 & 7.28 & 3.25    \\
\hline
\end{tabular}
\end{table}

The analysis of simultaneous events among stations yielded 42 single-pulse candidates coincident within time windows of 86.4$\,$ms or less. Further inspection of the time series around the detected pulses showed that these candidates were produced by baseline level jumps or residual time series artifacts. From the visual inspection of all single-pulse candidates with $\sigma_{\rm SP}>12$, no convincing real pulses were found.

Using Equation~$1$ in \citet{kkl+15}, and the system sensitivity parameters outlined in Table~\ref{tab:smin_popcoverage} 
we estimate $\sigma_{\rm SP}=6$ or $\sigma_{\rm SP}=10$ on source sensitivity limits to representative 1, 10 and $50\,{\rm ms}$ duration single pulses. These limits are outlined in Table~\ref{tab:spsearch_params_results}.  



\section{Discussion} \label{sec:discussion}


\subsection{Sensitivity of the periodicity searches} \label{sec:sys_sensi}

\begin{figure}
\centering
  \centering
  \includegraphics[scale=0.56]{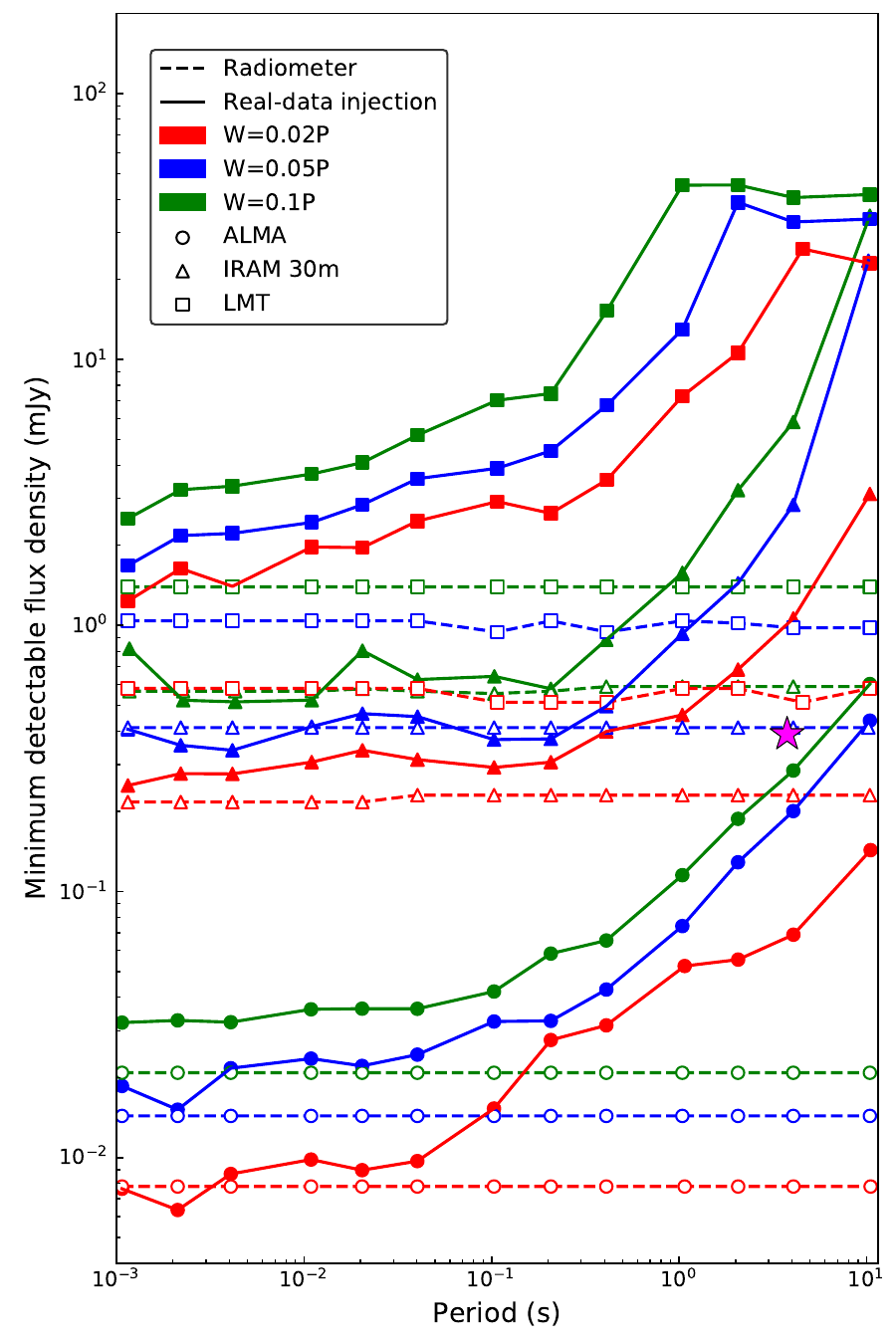}
  \caption{Theoretical (dashed lines) and empirical (solid lines) sensitivity estimate of the search on Apr. 7 (considered the most sensitive in all three epochs) at ALMA, IRAM~30\,m and LMT. The red, blue and green colour represent injection of signal with 2\%, 5\% and 10\% duty cycle, respectively. In the empirical estimate, each point represents the median from 5 injection iterations. The star in purple stands for the GC magnetar PSR~J1745$-$2900 using the flux density reported by \cite{lde+21} and assuming a flat spectrum. Here we used Equation~\ref{eq:radiometer_cc97} (with $\sigma=5$) to calculate the theoretical estimate of minimum detectable flux density.  
  }
  \label{fig:sensi}
\end{figure}

Several potential factors contribute to the non-detection of pulsars in this search,
including the typical steep spectrum of radio pulsars and the large distance to the center of our Galaxy. The sensitivity of a pulsar search is usually quantified by the minimum detectable mean flux density of a pulsar, $S_{\rm min}$, given a certain set of observational and pulsar properties. Based on the radiometer equation, the theoretical sensitivity of a Fourier-domain pulsar search can be expressed following \citep{cc97}
\begin{equation}
    \displaystyle S_{\rm min}= \frac{\beta \,\sigma \, T_{\rm sys}^* }{\eta_{\rm ph}\rm{G}\sqrt{\rm{n_p}\,\Delta\nu\,t_{\rm int}}} \left(\frac{1}{\alpha \sqrt{\rm{N_{\rm h}}}}\sum_{l=1}^{N_{\rm h}}R_{l}\right)^{-1},
    \label{eq:radiometer_cc97}
\end{equation}
where $\beta$ is a correction factor due to imperfections in digitization \citep[$\beta=1.136$ in our case for 2-bit sampling,][]{coo70}, $\eta_{\rm ph}$ is the phasing efficiency of a phased array \citep[$\eta_{\rm{ph}}=1$ for single-dish telescopes, and $\eta_{\rm{ph}}\simeq0.95$ for phased ALMA observations during the EHT 2017 observations,][]{goddi19}, $\sigma$ is the requested detection significance of the pulsar signal, $T_{\rm sys}^*$ stands for the system temperature (including the contribution due to Earth's atmopshere), G is the telescope gain, $\rm{n_p}$ is the number of summed polarizations, $\Delta\nu$ is the effective observing bandwidth, $t_{\rm int}$ is the integration time, $\alpha\equiv\sqrt{1-\pi/4}$ is a coefficient related to the RMS noise in the Fourier transform of the intensity, $\rm{N_{\rm h}}$ is the optimum number of harmonics summed\footnote{The $\rm{N_{\rm h}}$ parameter refers to the number of summed harmonics that minimizes $S_{\rm min}$. It depends on pulse duty cycle and shape \citep[see Appendix A in][for details]{cc97}. $\rm{N}_{\rm h}$ may not coincide with the maximum number of harmonics that the pipeline can sum. The {\sc presto} pipeline searches for the optimal number of summed harmonics for each detected periodicity up to the limit, which in our case is 32 harmonics.} and $R_l$ is the amplitude ratio of the $l$th harmonic of the signal to the zero frequency in the Fourier transform. Assuming a Gaussian profile shape of the pulsar signal with a width $w$ (the pulse width at half height of the pulse), $R_l$ can then be written as
\begin{equation}
    R_l=e^{-(\pi\epsilon l/2\sqrt{{\rm ln}\,2})^2},
\end{equation}
where $\epsilon\equiv w/P$ is the duty cycle of the pulse. Based on these, the theoretical sensitivities of our searches can be computed using the telescope and data properties summarised in Tables~\ref{tab:searchparams_and_candidates}, \ref{tab:cal_info_gamma} and \ref{tab:smin_popcoverage}. These results are shown in Figure~\ref{fig:sensi}.

\begin{table}
\centering
\caption{Calibration information per station and epoch. Remaining columns show  the telescope gain (G), the average effective system temperature (that includes the contribution due to Earth's atmosphere, $T_{\rm sys}^*$), the net integration time on \sgra ($t_{\rm int}$) and a figure-of-merit of the sensitivity of each dataset, $\Gamma={ 1000 \cdot T_{\rm sys}^* } / ( \eta_{\rm{ph}} \, \rm{G} \, \sqrt{ 2 \, t_{\rm{int}} \, {\Delta \nu} } )$.
}
\label{tab:cal_info_gamma}
\begin{tabular}{clcccc}
\hline
\hline
Station &  Epoch    & G                     & ${T_{\rm sys}}^*$ & $t_{\rm int}$ & $\Gamma$  \\
        &           & ($\rm{K\,Jy^{-1}}$)    & (K)               & (hr)          & ($\mu\rm{Jy}$)\\
\hline 
ALMA    & Apr.\,6    & 1.054 & 127   & 2.1   & 17    \\
        & Apr.\,7    & 1.054 & 121   & 4.6   & 11    \\
        & Apr.\,11   & 1.054 & 196   & 1.9   & 27    \\
LMT     & Apr.\,6    & 0.033 & 355   & 2.4   & 1294  \\
        & Apr.\,7    & 0.063 & 490   & 3.0   & 837   \\
        & Apr.\,11   & 0.067 & 436   & 1.7   & 930   \\
\mbox{IRAM$\,$30$\,$m} & Apr.\,7   & 0.123 & 223   & 1.1   & 322   \\
\hline 
\end{tabular}
\end{table}

In practice, Equation~\ref{eq:radiometer_cc97} is known to overestimate the realistic sensitivity of a pulsar search. This is particularly true for slow pulsars with large duty cycles, when red noise is prominent in the data \citep{lbh+15,lde+21,etd+21}. Therefore, a more practical accurate estimate 
is obtained empirically by injecting synthetic pulsar signals each time of a different flux density into the real data, conduct the search and obtain the minimum detectable flux density.
Such a procedure has been carried out using the longest and most sensitive (see Table~\ref{tab:cal_info_gamma}) dataset from each of the three telescopes (all on Apr. 7) and the results are presented in Figure~\ref{fig:sensi}. It can be seen that for fast spinning pulsars, the estimate from injection into real data gives a minimum detectable flux density close to the theoretical estimate, compared with those for slow pulsars. As the pulsar period increases, e.g., from 1\,ms to 1\,s, the empirical sensitivity of the search drops drastically, by up to more than an order of magnitude.

The most sensitive search comes as expected from phased ALMA, where $S_{\rm min}\sim0.01\,{\rm mJy}$. For the single-dish telescopes, where the field of view is significantly larger, the best sensitivity is yielded by the \mbox{IRAM$\,$30$\,$m} telescope, 
where $S_{\rm min}\sim0.4\,{\rm mJy}$.
For the LMT, $S_{\rm min}\sim 1\,{\rm mJy}$. The GC magnetar PSR~J1745$-$2900 had a flux density of 0.39\,mJy at 86\,GHz 
on Apr. 3, 2017; the closest published observation, in both epoch and frequency, to those of the EHT \citep{lde+21}. Under the assumption of
a flat spectrum \citep[e.g.,][]{tek+15,tde+17}, PSR~J1745$-$2900 would fall below the empirically-derived sensitivity limits of \mbox{IRAM$\,$30$\,$m} and LMT and would explain our non-detection in the data from these two telescopes (see Figure~\ref{fig:sensi}). Although phased ALMA may have had enough sensitivity to detect PSR~J1745$-$2900 during the EHT 2017 observations, its field of view of 1--2\,\arcsec was not wide enough to cover this source.

The sensitivity estimates given above are primarily for the Fourier-domain search methods. The sensitivity of the FFA is in theory on a similar level. In \citet{2020MNRAS.497.4654M}, the theoretical sensitivity difference between these two methods is characterized by the search efficiency factor ($\mathcal{E}$), a function of the number of harmonics summed and duty cycle of the pulsar. With a summation of up to 32 harmonics, this factor is approximately 0.65, 0.76, 0.83 for signals with duty cycle of 2\%, 5\%, 10\% (used in our estimates above), respectively. Assuming $\mathcal{E}_{\rm FFA}=0.93$ as in \citet{2020MNRAS.497.4654M}, the theoretical sensitivity of the FFA search is better than the FFT search by a factor of 1.4, 1.2, 1.1, respectively. 



Finally, to illustrate the advantage of our search at very high frequency to overcome the scattering effect, we consider a pessimistic (but still possible) scenario of temporal scattering toward the Galactic Center as predicted by the ${\rm NE}2001$ model, i.e., $\tau_{\rm s} = 2000\, \nu^{-4}\,$s ($\nu$ in unit of GHz), where $\tau_{\rm s}$ is the exponential characteristic time on the pulses due to the scattering, and $\nu$ is the observing frequency. Even in this case, at 228\,GHz the scattering time would be $\tau_{\rm s} \approx 740\,\rm ns$, roughly 40 times smaller than our smallest sampling interval, three orders of magnitude less than the spin period of any pulsar known, and $\sim50$ times less than the narrowest pulse width known to date \citep{mhth05}. We note that the scattering measured for the currently-known closest pulsar to \mbox{Sgr$\,$A*}, PSR~J1745$-$2900 \citep[$\tau_{\rm s}\simeq1.3\nu^{-3.8}$;][]{spi14}, is much smaller than the predicted value from the NE2001 model. However, we cannot fully discard a patchy environment in the innermost region of the Milky Way, with different scattering for different line-of-sights due to multiple screens \citep[e.g.,][]{schni16, dex2017GCscatt}. In any case, due to the very-high observing frequency of the EHT, our search sensitivity is unaffected by interstellar scattering.

\subsection{Potential of the search to detect a Galactic Center pulsar population}\label{ssec:search_deepness}

\begin{table*}
	\centering
	\caption{Parameters from each station, corresponding to the most sensitive datasets, used for the sensitivity analysis. Columns indicate the station, observing frequency ($\nu$), average system temperature ($T_{\rm sys}^*$), telescope gain (G), net integration time on-source ($t_{\rm int}$), instantaneous bandwidth ($\Delta{\rm \nu}$), assumed pulsar duty cycle for the calculations ($\epsilon$), resulting theoretical minimum detectable flux density for $\sigma=5$ ($S_{\rm min}$), equivalent pseudo-luminosity at the distance of the Galactic Center ($L_{\rm min}^{\rm GC}$), and the percentage of the simulated pulsar population in the Galactic Center that the observations could detect for the theoretically- ($f_{\rm t}$) and empirically-derived ($f_{\rm e}$) luminosity limits (see Section~\ref{ssec:search_deepness}). In all cases the number of summed polarization $\rm{n_p}=2$. The two last rows show the results from similar recent pulsar searches around \mbox{Sgr$\,$A*} at wavelengths of 3.5$\,$mm with ALMA \citep{lde+21} and 3 to 2$\,$mm with the \mbox{IRAM$\,$30$\,$m} telescope \citep{tor21}, for comparison. The ``n/a'' abbreviation indicates that a result is not available.}
	\label{tab:smin_popcoverage}
	\begin{tabular}{lcccccccccc} 
		\hline
		\hline
		Station & $\nu$ & $T_{\rm sys}^*$ & G & $t_{\rm int}$ & $\Delta\nu$ & $\epsilon$ & $S_{\rm min}$ & $L_{\rm min}^{\rm GC}$ & Detected $f_{\rm t}$ & Detected $f_{\rm e}$ \\
		        & (GHz) & (K)           & ($\rm{K\,Jy^{-1}}$)& (h) & (GHz) & & (mJy) & ($\rm mJy\,kpc^2$) & (\%) & (\%) \\
		
		\hline
		ALMA      & 228  & 121  & 1.054 &  4.6 & 3.75 & 0.05  & 0.01  & 0.7   & 4.1  & 2.2 \\
		LMT       & 228  & 490  & 0.063 &  3.0 & 4.0  & 0.05  & 1.09  & 75   & 0.3  & 0.1 \\
		\mbox{IRAM$\,$30$\,$m} & 228 & 223  & 0.123 &  1.1 & 4.0  & 0.05  & 0.42  & 29    & 0.4  & 0.4 \\
        
        \rule{0pt}{3ex}ALMA    & 86    & 51  & 1.15 & 3.0 & 2.0 & 0.05  & 0.01  & 0.7   & n/a  & 4.0 \\
        \mbox{IRAM$\,$30$\,$m} & 86-154 & 125  & 0.16 &  3.0 & 32  & 0.1  & 0.06  & 4.0    & 1-2  & n/a \\
	
		\hline
	\end{tabular}
\end{table*}

Once the flux density limits of the observations are known, we can estimate the potential to detect pulsars around Sgr$\,$A* with this search. We do this by calculating how many pulsars, from a simulated population in the GC, would be detectable with our actual sensitivity. In the following analysis we focus on the most-sensitive observation from each station as representative of the best limit for GC pulsar detectability, i.e., the data from April 7 (see Table~\ref{tab:cal_info_gamma}).

The flux density limit from Equation~\ref{eq:radiometer_cc97} or through the injection of signals (Section~\ref{sec:sys_sensi}) can be converted to a pseudo-luminosity limit at the GC by multiplying $S_{\rm {min}}$ by the square of the distance to the GC \citep[see e.g.,][]{lorkra04}, $d_{\rm GC}=8.28\,\rm kpc$ \citep{gaa+21}:
\begin{equation}\label{eq:lumthres}
    L_{\rm min}^{\rm GC} = S_{\rm min}\,d_{\rm GC}^2 = S_{\rm min}\,8.28^2 \;\;\rm [mJy\,kpc^2].
\end{equation}
The simulated population in the GC is created from the known pulsar population in the Milky Way, i.e., we assume that the putative pulsars at the center of the Galaxy resemble in properties those of the Milky Way. Although the properties of the pulsars that may populate the innermost region of the Galaxy are unknown, we consider this is a valid assumption given that both old and new populations of stars exist in the GC \citep{pfuhl2011_GCstars, nl22}. We use the online pulsar catalog PSRCAT\footnote{\url{https://www.atnf.csiro.au/research/pulsar/psrcat/}} \citep{mhth05} in its version 1.67,
extracting the pseudo-luminosity of those pulsars for which this value is available (in this catalog version, from 3320 pulsars, 2457, i.e. 74\%, have a pseudo-luminosity entry). These pseudo-luminosity values are in most cases provided for frequencies of 400 and 1400$\,$MHz. We therefore extrapolate to 228$\,$GHz by using the spectral index of each pulsar that is available in the catalog. When the spectral index is not available, we draw a sample from a normal distribution with mean value $\left<\alpha_1\right>=-1.60\pm0.54$ \citep{jank18}.
Both when the spectral index is known and when we draw from the distribution, a single power law for the extrapolation is used. Finally, we compute the percentage of pulsars with an extrapolated pseudo-luminosity at 228$\,$GHz above the pseudo-luminosity limit obtained with Equation~\ref{eq:lumthres}, both with the theoretically- and empirically-derived $S_{\min}$. Those percentages of the simulated pulsars would be in theory detectable in this search. 

\begin{figure*}
\centering
  \centering
  \includegraphics[width=\textwidth]{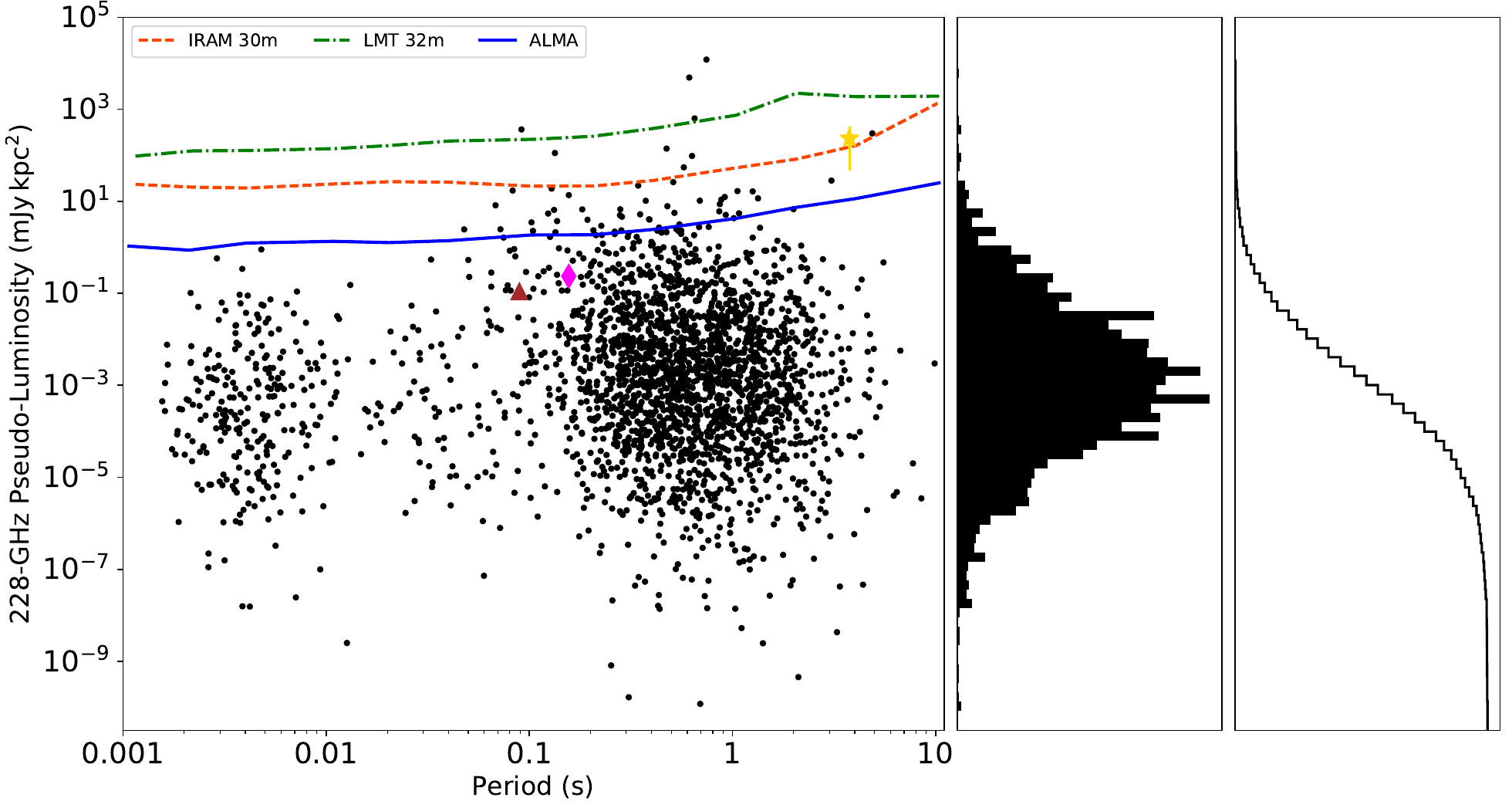}
  \caption{(Left:) Luminosity for the simulated pulsar population at 228$\,$GHz (black dots) with the empirically-derived minimum detectable luminosity for each telescope at the distance to the Galactic Center ($d=8.28\,\rm kpc$) over-plotted as lines. The minimum detectable luminosity corresponds to the empirical limits on flux density as seen in Figure~\ref{fig:sensi} for a pulse duty cycle of 5\%. The blue line, dashed-dotted green line, and dashed orange line represents the limits for ALMA, LMT and the \mbox{IRAM$\,$30$\,$m} telescope, respectively. Each black dot represents a simulated pulsar, and the dots above each line would in theory be detectable for that telescope. For reference, PSR~J1745$-$2900's average luminosity at 228$\,$GHz \citep{tek+15, tde+17}, the Vela pulsar (PSR~B0833$-$45) and PSR~B0355+54 are marked with a yellow star, brown triangle and pink diamond symbols, respectively. The vertical line over the yellow star marks the range of known variability of PSR~J1745$-$2900. (Middle:) Histogram of the luminosity distribution, and (Right:) histogram of the luminosity distribution in a cumulative layout.}
  \label{fig:lumplot}
\end{figure*}

Because the exact number of pulsars above the limit will depend on the samples drawn from the normal distribution of spectral indices, we run a Monte Carlo simulation with 10000 executions to obtain the average percentage of detectable population for each station. The summary of the utilized parameters and the results are presented in Table~\ref{tab:smin_popcoverage}. Figure~\ref{fig:lumplot} shows a snapshot of the Monte Carlo simulation (representative of the results), together with the empirical luminosity limits for each station. 

In addition, to investigate how the spectral index distribution could affect the results, we repeated the analysis in other scenarios for the simulated pulsar population: one with $\left<\alpha_2\right>=-1.4\pm1.0$ \citep{bates2013}, and with $\left<\alpha_3\right>=-1.8\pm0.2$ \citep{mar2000}. Similarly, the impact of the duty cycle parameter was evaluated by repeating the simulation, only for the spectral index distribution $\left<\alpha_1\right>$, for $\epsilon=0.01$, and $0.2$.

Focusing on the results from the empirical sensitivity, the spectral index distribution $\left<\alpha_2\right>$ results in a detectable percentage of the simulated population of 8.8\%, 1.3\%, and 3.4\% for ALMA, LMT and the \mbox{IRAM$\,$30$\,$m} telescope, respectively. In the case of using $\left<\alpha_3\right>$, the percentage decreases to 1.4\%, 0.1\%, and 0.3\%, respectively. These results show that the percentage of detectable population can vary a factor $\sim$2 to 10 depending on the spectral index distribution utilized; mainly a result of the required extrapolation to 228$\,$GHz and the difference in the standard deviation of the assumed distribution. The simulations show that the mean value itself has a smaller impact on the result than the standard deviation. Even with a steeper mean value, if the standard deviation is sufficiently large, pulsars with flat spectral indices and thus more likely emitting above our sensitivity threshold at 228$\,$GHz, will be drawn from the distribution. In the future, understanding better the distribution of pulsar spectral indices at high radio frequencies ($\nu\gtrsim20-30\,$GHz) would help to obtain more accurate estimations for the potentially-detectable fraction from pulsar searches at very high radio frequencies \citep[see][]{lohmer08}. In contrast, the simulations with additional values of the duty cycle $\epsilon=0.01$ and $0.2$ in the $\left<\alpha_1\right>$ scenario shows negligible variations on the detectable population.

A first conclusion from the results summarised in Table~\ref{tab:smin_popcoverage} is that our potential to detect pulsars in the Galactic Center with this search is in general low. This was somewhat expected, primarily due to the effect of the average steep spectral index of pulsars. However, with the observations at 228$\,$GHz we were mainly targeting flat-spectrum pulsars, like radio magnetars, and probing a frequency space where certain emission models include possible emission boosts, making certain pulsars potentially visible. Furthermore, the single pulse analysis may detect pulsars through bright single pulses. 

A second finding shows that the percentage of detected pulsars can decrease by up to $\sim50-60\%$ when using the empirically-derived sensitivity limits versus the theoretical ones. This sensitivity loss is substantial, and is caused by the noise characteristics, which are far from showing the Gaussian statistics that the radiometer equation assumes. From the three stations, the data from the IRAM~30m seems to be the less affected by this loss of sensitivity. This is apparent in Figure~\ref{fig:sensi}, where the empirical limit for this station is flatter, and starts rising at longer spin periods as compared to the ALMA and LMT cases.

The highest fraction of detectable population in the EHT 2017 data comes from ALMA that, despite its superb sensitivity, could detect only around 2.2\% of the brightest pulsars in the simulated population with $\left<\alpha_1\right>$ --- that we consider as the most accurate. This fraction is lower by a factor $\sim$2 compared to the one yielded by ALMA in 2017 Global Millimeter VLBI observations (GMVA), of about $4\%$, despite the use of only 2$\,$GHz of bandwidth \citep{lde+21}.  
In the case of the \mbox{IRAM$\,$30$\,$m} telescope and the LMT, less than 0.4\% and 0.1\% of the simulated pulsars, respectively, would be detectable with the EHT 2017 data. For the \mbox{IRAM$\,$30$\,$m}, previous searches at 3 and 2$\,$mm yielded theoretical population coverage fractions of $1-2\%$ \citep{tor21}. The differences arise from a combination of different search parameters (e.g., ${T_{\rm sys}}^*$, G, $t_{\rm int}$, or $\Delta\nu$), together with the higher average luminosity of pulsars at the longer wavelengths.

Focusing on the EHT 2017 data, the simulations show therefore that only a very small number of very bright pulsars at 228$\,$GHz (1.3$\,$mm), i.e., those that not only are bright but also show a flat (or inverted) spectrum, had chances to be detected in this search. In fact, even bright pulsars at cm wavelengths like Vela, and bright, relatively flat-spectrum pulsars like PSR~B0355+54, would not be detected in our survey if located at the Galactic Center. Consequently, it is not surprising that no detections of new pulsars were achieved by our search. Related to our low sensitivity, it is therefore plausible that pulsars emitting below our current luminosity limits exist in the region. This is particular relevant for MSPs, that may be less luminous than normal pulsars \citep[e.g.,][]{bai97, kra98}. From our simulations, our sensitivity to MSPs in this search is close to zero (see Figure~\ref{fig:lumplot}). Given that the GC may be dominated by an MSP population \citep{gormac13, macq15, schoedel20}, this could also explain the lack of discoveries.

We remark that even pulsars emitting above our detection limits and with orbital parameters within the searching pipelines capabilities (see Section~\ref{ssec:sensi_orb}) may be missed by other factors. For example, a pulsar could be highly variable in flux density \citep[as occurs with radio-emitting magnetars, see e.g.,][]{tde+17}, or eclipses due to surrounding gas could hide a pulsar emission for some periods of time in binary systems \citep[e.g.,][]{freire2005}. Relativistic precession can move the beam of emission of a pulsar orbiting a companion in and out our line-of-sight \citep[e.g.,][]{kram98_prec, perera2010}. These reasons alone justify repeated surveys for pulsars around Sgr$\,$A* at different epochs separated in time to overcome these time-dependent effects. 

Lastly, another case potentially affecting our sensitivity is when a pulsar's spin and orbital parameters are beyond the capabilities of our pipelines to recover the accelerated signals. This mostly affects fast MSPs ($P\lesssim5\,$ms) in tight orbits around massive companions. Nevertheless, the possibility that the center of the Galaxy is populated to a large 
extent by MSPs cannot be ruled out \citep[e.g.,][]{rajwade2017}. Since the main signals aimed for here are pulsars orbiting the super massive black hole Sgr$\,$A*, we should consider our sensitivity to such MSP-(SM)BH systems potentially lower than the limits shown in Table~\ref{tab:smin_popcoverage} and Figures~\ref{fig:sensi} and \ref{fig:lumplot}. In the next Section, we discuss this relationship between sensitivity and binary systems in more detail.

\subsection{Search capability for pulsar in various orbits around \mbox{Sgr$\,$A*}} \label{ssec:sensi_orb}

The phased ALMA field of view is significantly smaller than LMT and \mbox{IRAM$\,$30$\,$m}, sampling a smaller volume and range of potential orbits near \mbox{Sgr$\,$A*}. Using the distance and mass of \mbox{Sgr$\,$A*} derived from \cite{gaa+21} and assuming a circular orbit, the phased ALMA field of view is able to cover an orbital period of approximately up to $P_{\rm b}\approx300\,$yr. This is already enough to cover a large fraction of the stars known to date in the S-star cluster \citep{sem+12, guill17}, and certainly also the orbits that would enable Sgr$\,$A* measurements and gravity tests with pulsars \citep{lwk+12}. Meanwhile, the field of view of LMT and \mbox{IRAM~30\,m} can cover orbital periods of up to $P_{\rm b}\approx4000\,$yr. So, even though their observations are less sensitive, the LMT and IRAM~30\,m telescope are able to cover a significantly larger volume of any putative pulsar population in the GC.

\begin{figure*}
     \setlength\tabcolsep{0pt}
     \centering
     \begin{tabular}{cccc}
         \includegraphics[width=0.25\textwidth]{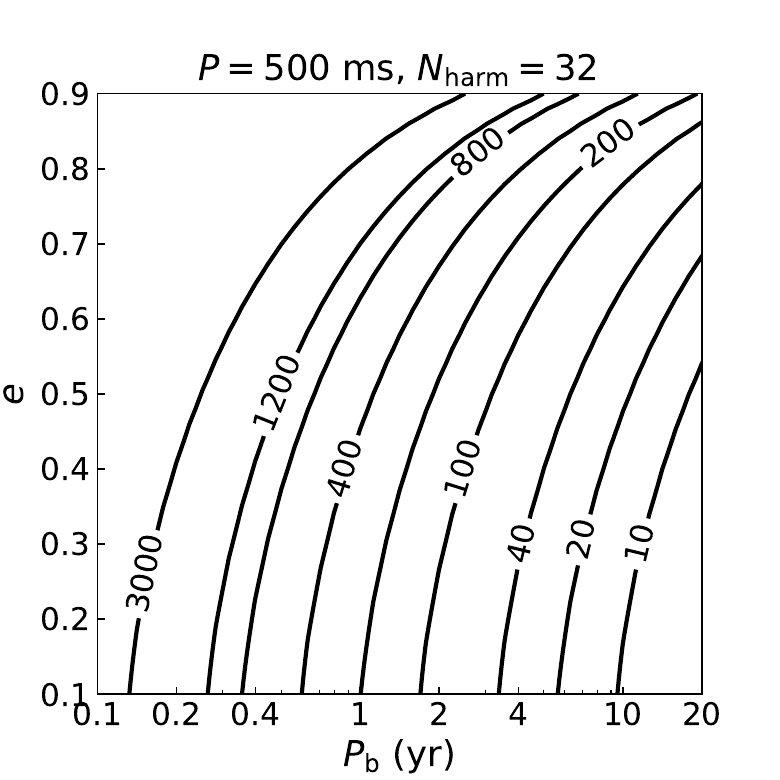} &
         \includegraphics[width=0.25\textwidth]{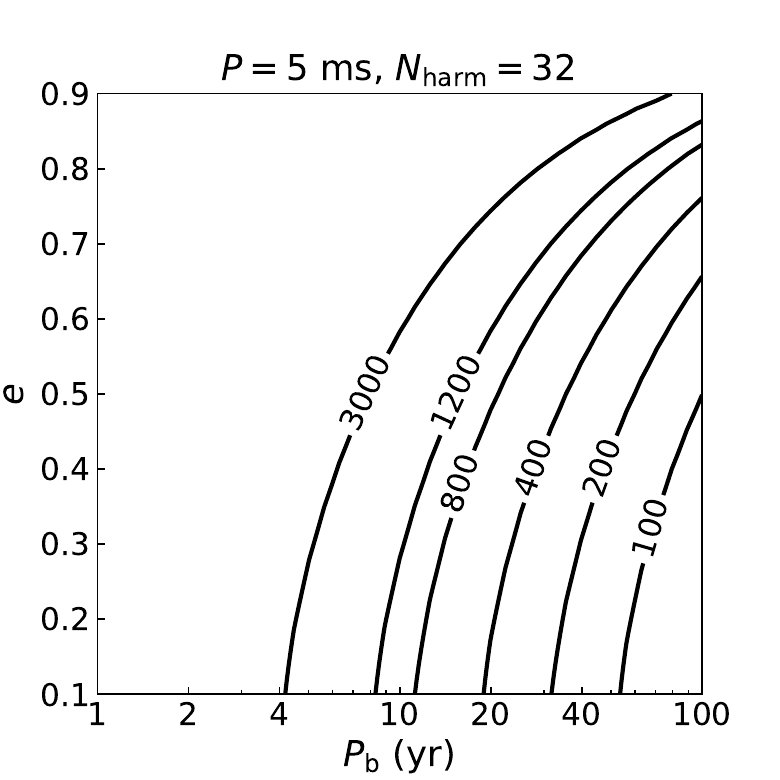} &
         \includegraphics[width=0.25\textwidth]{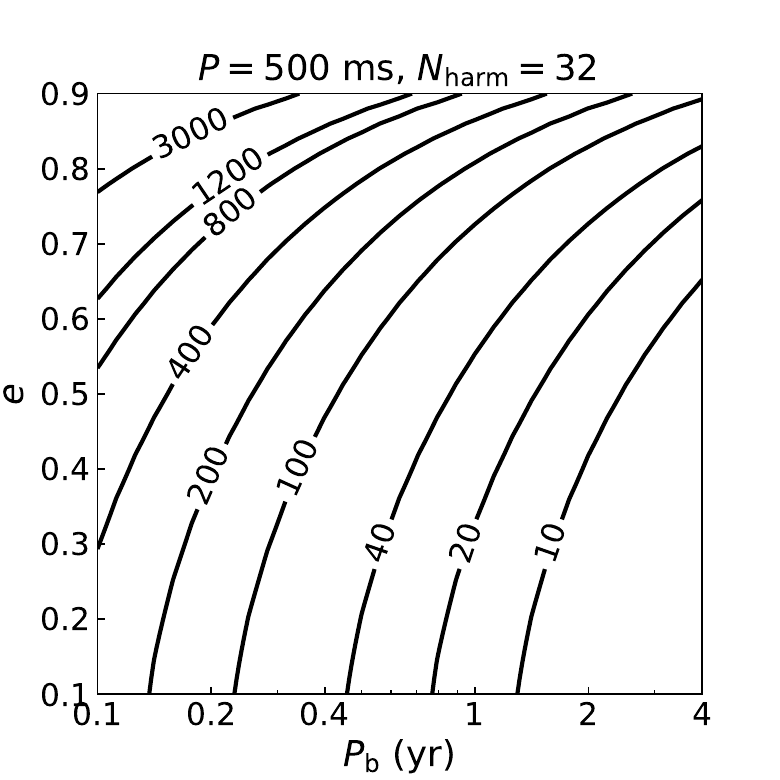} &
         \includegraphics[width=0.25\textwidth]{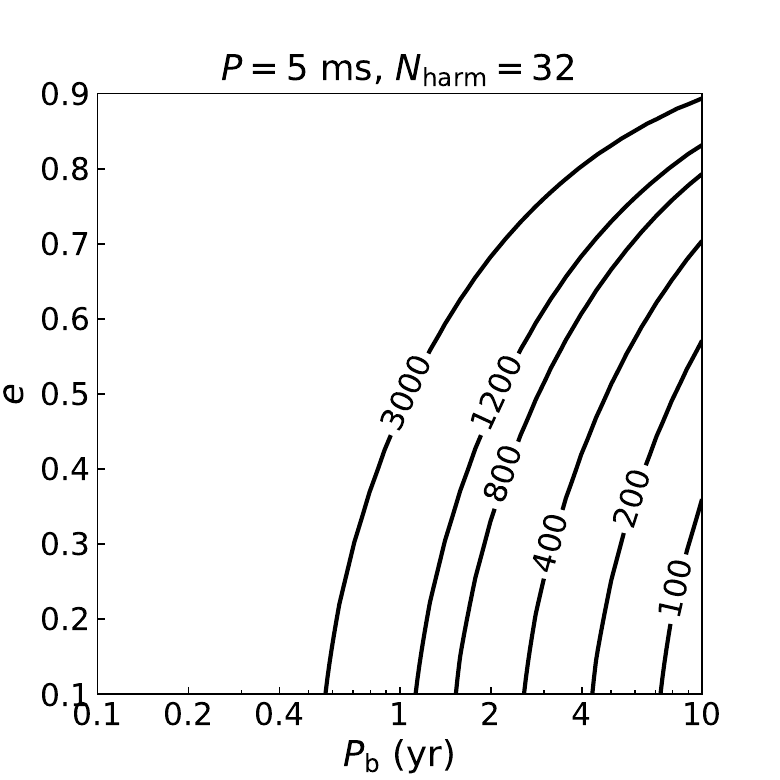} \\
         
         \includegraphics[width=0.25\textwidth]{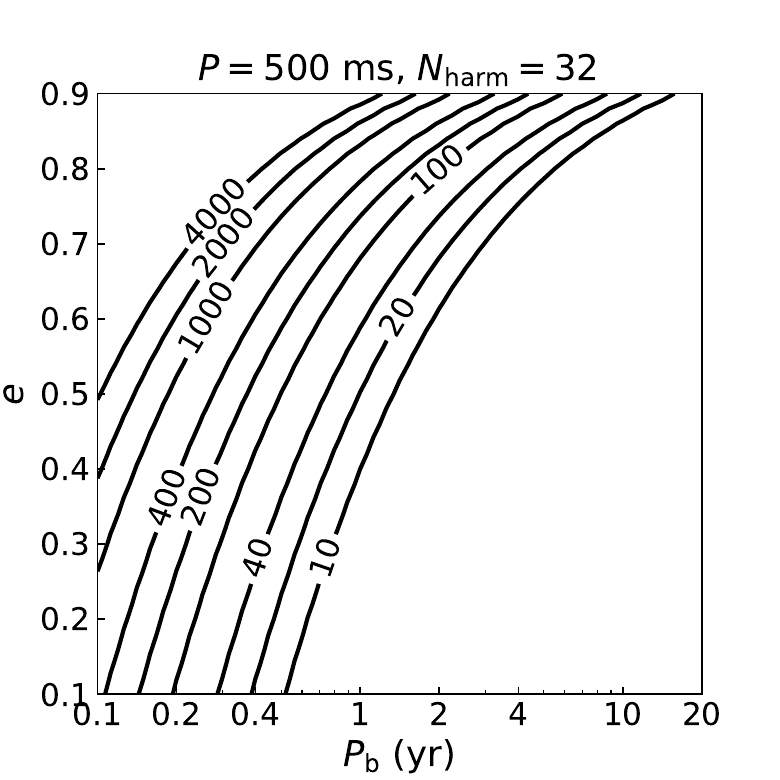} &
         \includegraphics[width=0.25\textwidth]{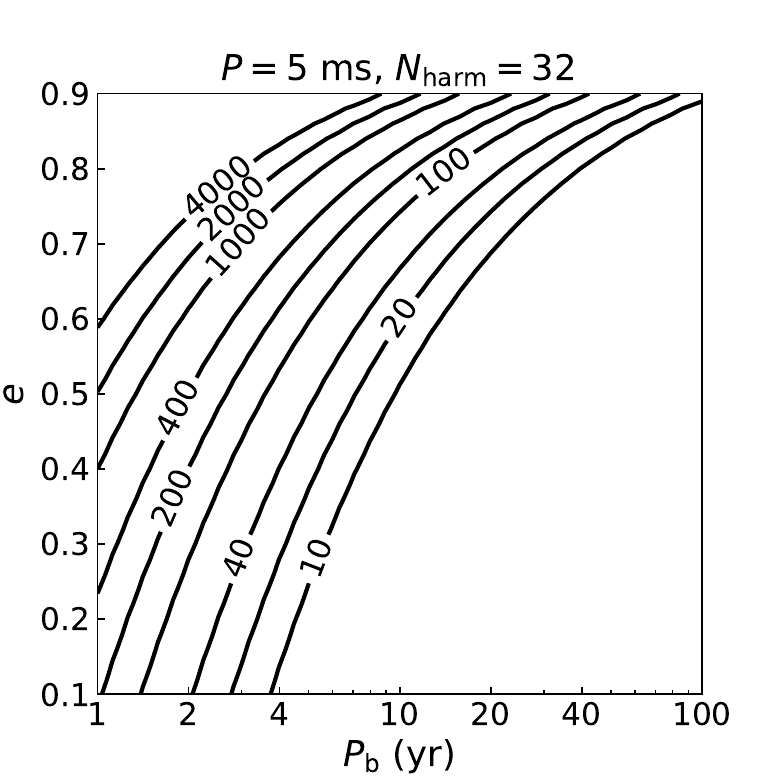} &
         \includegraphics[width=0.25\textwidth]{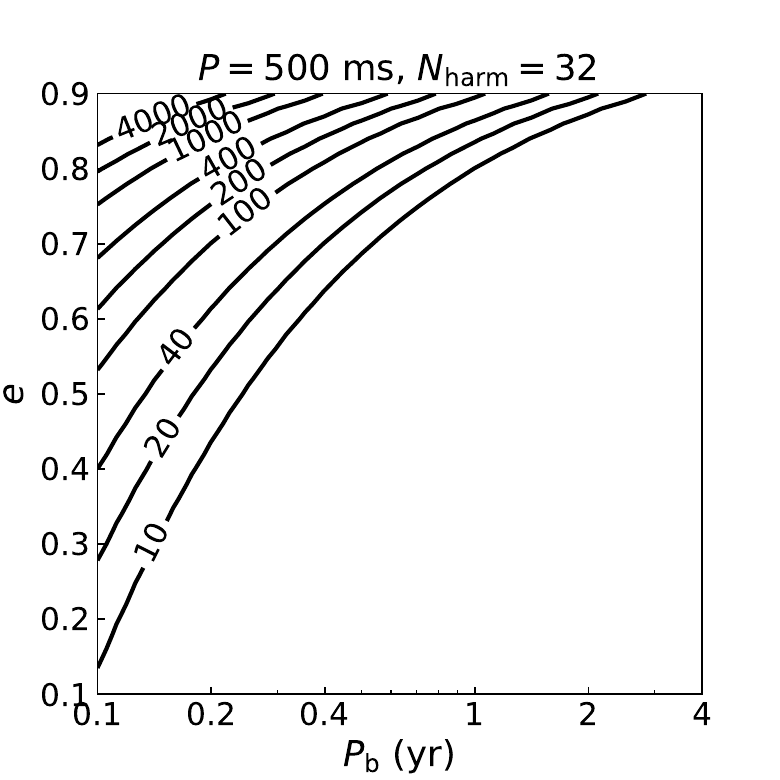} &
         \includegraphics[width=0.25\textwidth]{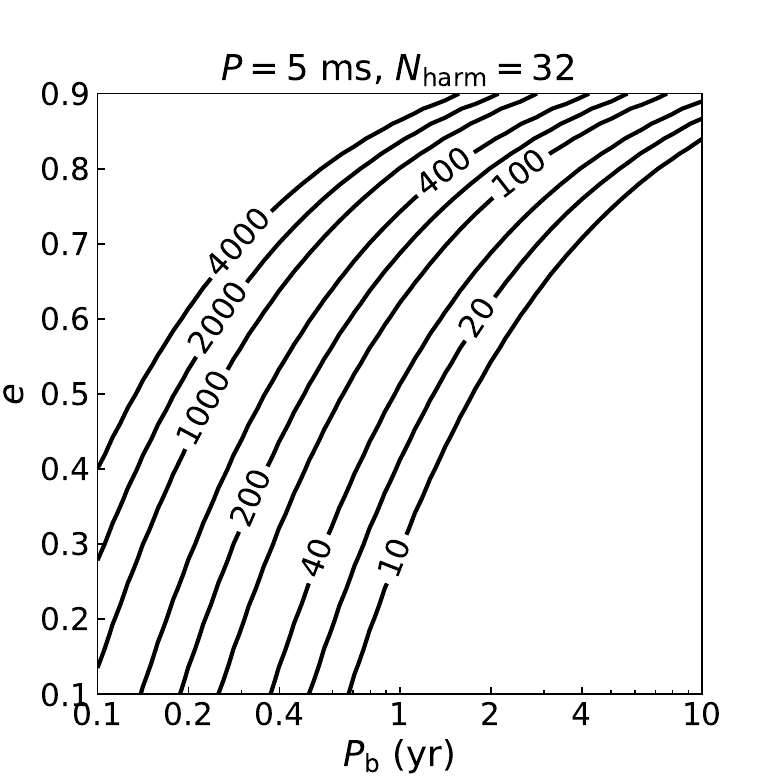} \\
         
         {(a) $T_{\rm obs}=10.2\,$hr} &
         {(b) $T_{\rm obs}=10.2\,$hr} &
         {(c) $T_{\rm obs}=3.2\,$hr} &
         {(d) $T_{\rm obs}=3.2\,$hr}
     \end{tabular}
     \caption{Values of \texttt{zmax} (upper row) and \texttt{wmax} (lower row) required for the Fourier-domain search to be conducted with optimal sensitivity in orbits around \mbox{Sgr$\,$A*} with different periods and eccentricity ($e$). Here we use two pulsar spin periods, which are typical values for ordinary pulsars and MSPs, respectively. The time spans of the observations are assumed to be 10.2 and 3.2$\,$hr, which are in turn the longest and shortest in our analyzed datasets. \label{fig:zwmax}}
\end{figure*}

In practice, as already mentioned in Section~\ref{sec:searches}, the range of acceleration and its derivative explored in the search would also place a constraint on the types of orbits around \mbox{Sgr$\,$A*} that the search is able to detect. For a given time span of the observation, the constraint can be estimated by calculating the maximum absolute $z$ and $w$ values in a range of orbits and compare with those used in the search, as demonstrated in \cite{lde+21} \citep[see also][]{etd+21}. Here as case studies, we carried out the same practice for two time spans of the observation, $T_{\rm obs}=10.2$ and $3.2$\,hr, which correspond to the April-7 observation of ALMA and \mbox{IRAM$\,$30$\,$m}, respectively, being the longest and shortest time span in our dataset. We focus on a full sensitivity signal recovery based on 32 summed harmonics (see Section~\ref{sec:FFTsearch}). The results are summarised in Figure~\ref{fig:zwmax}, with a selection of two typical spin periods for ordinary pulsars ($P\simeq500$\,ms) and MSPs ($P\simeq5$\,ms), respectively.

With the settings detailed in Section~\ref{sec:FFTsearch}, our search using 10.2-hr long data is capable of detecting ordinary pulsars well within one-year orbits around \mbox{Sgr$\,$A*}, likely down to $P_{\rm b}\gtrsim0.5$\,yr. For MSPs, we are able to detect those with an orbital period longer than 10\,yr. Meanwhile, the search using 3.2-hr long data can detect ordinary pulsars in much closer orbits, i.e., down to 0.2\,yr ($\approx2.4\,$months). For MSPs, it can detect orbits with as short as approximately 2-yr periods. These are also inferred by Equation~\ref{eq:zmax} and \ref{eq:wmax}, where the parameter space in acceleration and jerk to be explored scales down for short observations and slow pulsars. Note that as shown in Table~\ref{tab:observations_summary}, the ALMA observations from April 6 and 11 have similar time span and thus coverage of pulsar orbits as the \mbox{IRAM$\,$30$\,$m} observation.

As discussed in \cite{lde+21}, the above limit estimate on the orbits corresponds to the boundary where the optimal sensitivity can always be retained throughout the entire orbit. If in practice the search conducted does not cover the recorded ${\rm max}(|z|)$ and ${\rm max}(|w|)$, it would start to lose sensitivity in the worse scenario but could still detect the pulsar. For instance, for highly eccentric orbits, the pulsar spends most of the time 
near the apoastron, where the acceleration and its derivative are considerably lower than those near periastron. In this case, searches that do not cover ${\rm max}(|z|)$ and ${\rm max}(|w|)$ of the entire orbit, could still be able to find the pulsar at a large fraction of the orbital phase.


\subsection{Constraints on a putative Galactic Center pulsar population} \label{sec:GCpsr_constrains}

Theoretical estimations on the number of pulsars that may exist in the inner part of our Galaxy differ considerably, with typical assumed values from around $\sim$200 \citep{chen14} to thousands \citep{whar12}. The wide range relates mainly to the complexity and peculiarity of the star forming history and evolution in the Galactic Center, and the dynamics that the high-density environment and the SMBH introduce \citep[see e.g.,][]{figer09, morris23}. Using our survey results, we can try to constrain the number of pulsars that may exist in the Galactic Center. 

We will assume a population of pulsars located at the distance of \sgra. For simplicity in the calculations, the population will 
have a Gaussian radial distribution, with two parameters: $n_0$ as the number density of pulsars, and $\sigma_{\rm R}$ the RMS radial scale, for a 3D Gaussian distribution. Following a similar approach as in \citet{cc97}, integrating over the angles sampled by each telescope we calculate the mean expected number of pulsar detections from this assumed population ($\rm{N}_{\rm d}$), for a given telescope, as

\begin{equation}\label{eq:Nd}
    {\rm N}_{\rm d}(n_0, \sigma_{\rm R}) = (2\pi)^{3/2}\,n_0\,\left[\frac{f_{\rm s}\,{\sigma_{\rm R}}^3}{1+{\sigma_{\rm R}}^2 / (\rm{d_{gc}} \sigma_\theta)^2}\right],
\end{equation}

where $f_{\rm s}$ is the fraction of detectable pulsars with a given telescope (we use the empirical detection fraction $f_{\rm e}$ for our calculations, see Table~\ref{tab:smin_popcoverage}), $\rm{d_{gc}}$ is the distance to the Galactic Center, and $\sigma_\theta$ is the RMS angular scale of a Gaussian beam pattern for the telescope.

\begin{figure*}
\centering
  \centering
  \includegraphics[width=0.95\linewidth]{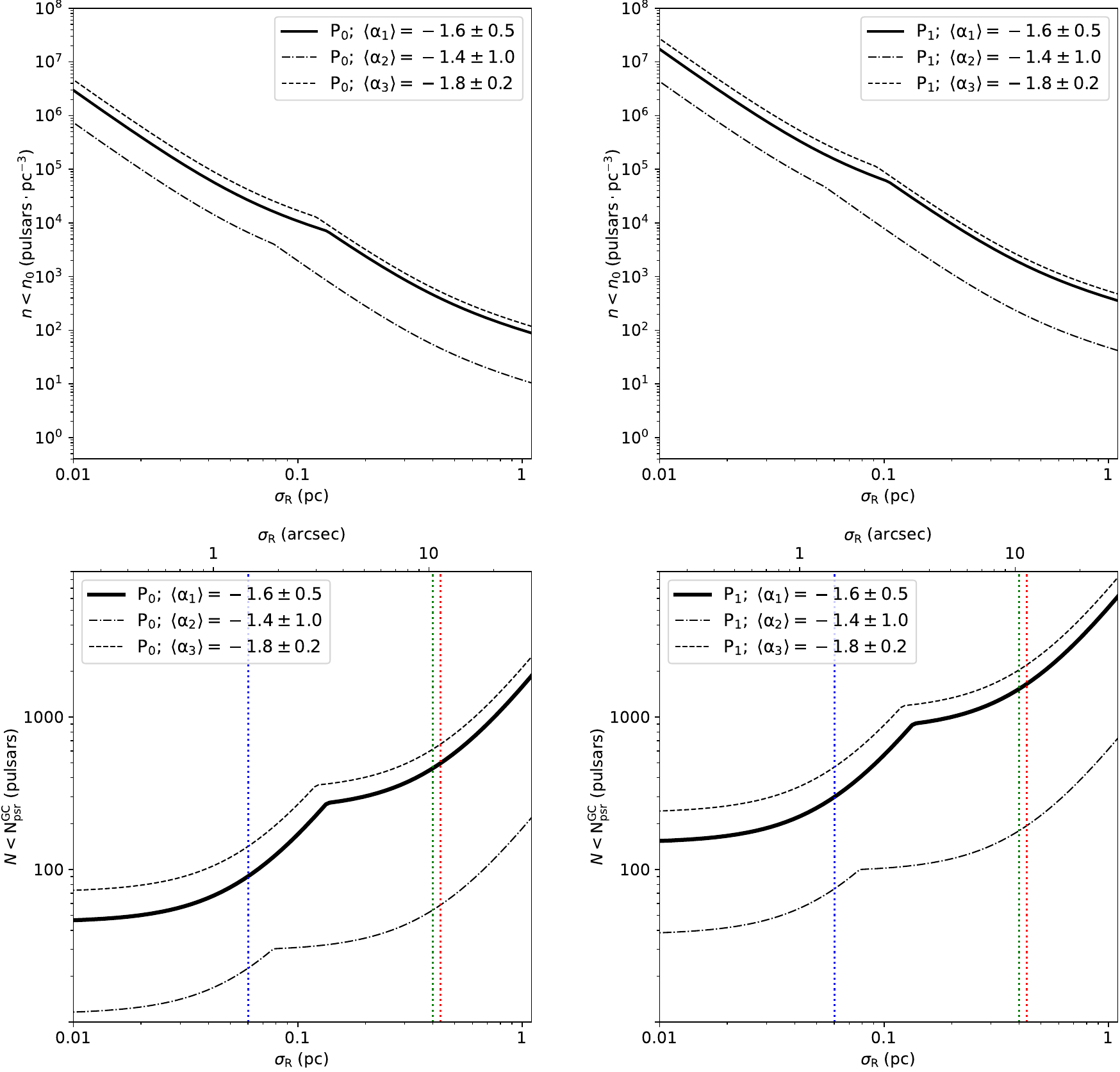}
  \caption{(Top panels:) Upper limits on pulsar number density ($n_0$) versus Gaussian radial width ($\sigma_{\rm R}$) constrained by the EHT 2017 observations for a survey likelihood $\mathcal{L}_0 = \rm{P_0}$ (left) and $\mathcal{L}_1 = \rm{P_1}$ (right). (Bottom panels:) Upper limit on the number of expected pulsars around \sgra within a spherical Gaussian volume with radial width $\sigma_{\rm R}$. The top axis indicates the Gaussian radial width in arcsec scale, for easier comparison with the telescope beam sizes, represented by the vertical dotted lines in blue, green, and red color, at $\approx$0.06, $\approx$0.4 and $\approx$0.43$\,$pc for phased ALMA, LMT, and the IRAM~30m, respectively. The survey likelihood function is $\mathcal{L}_0$ (left) and $\mathcal{L}_1$ (right). In all the panels the continuous thick, dotted-dashed and dashed lines marks the upper limits for a pulsar population with a mean spectral index $\left<\alpha_1\right>$, $\left<\alpha_2\right>$ and $\left<\alpha_3\right>$, respectively.}
  \label{fig:Npsr_GC_Combined}
\end{figure*}

To constrain the parameters of the population, we assume that our probability for detections follows a Poisson distribution. Given our non-detections, we build our survey likelihood function as $\mathcal{L}_0 = \rm{P_0} = e^{-\rm{N}_{\rm d}}$. We note that the two parameters for the assumed GC pulsar population are not statistically independent, so the upper limits for $n_0$ will depend on $\sigma_{\rm R}$. We calculate an upper limit for $\rm{N}_{\rm d}$ (and so, also for $n_0$) at a given confidence level, that we choose here as 99.7\%. In order to obtain the most stringent constraint on the expected number of pulsars, we select the lowest $n_0$ for each $\sigma_{\rm R}$ from the values obtained from the three used telescopes. 
Finally, we compute an upper limit for the expected mean number of total pulsars in the assumed population as ${\rm N}_{\rm psr}^{\rm GC}=n_0\,\rm{V}_{\rm pop}$, where $\rm{V}_{\rm pop}=(2\pi)^{3/2}\,\sigma_{\rm R}^3$ is the volume enclosing the assumed population. The results are plotted in Figure~\ref{fig:Npsr_GC_Combined}.

Although we have not detected any pulsar in the EHT 2017 observations, we know that at least one pulsar exist near \sgra: the magnetar PSR~J1745$-$2900, just three arcsec away \citep[i.e., a projected distance of $\approx0.1\,{\rm pc}$,][]{rea13}. This pulsar was not detected because it is highly variable and was likely on a weak-emission state during the EHT 2017 campaign. PSR~J1745$-$2900 would have been detected by the IRAM 30m telescope, and even phased ALMA (even offset from the center of its synthetic beam), if it were in a bright-emission episode \citep[PSR~J1745$-$2900 can reach $S\sim6\,$mJy at 1.3$\,$mm,][]{tde+17}. Taking into account an alternative scenario with a detection of PSR~J1745$-$2900, we also present the upper limits considering our survey likelihood function $\mathcal{L}_{1} = \rm{P_1} = \rm{N}_{\rm d}\,e^{-\rm{N}_{\rm d}}$ in Figure~\ref{fig:Npsr_GC_Combined}.

The derived upper limits show that, given the assumptions made on pulsar population properties (i.e., similar in properties to the known pulsar population, and with the chosen spectral index distribution $\left<\alpha_1\right>=-1.60\pm0.54$), and the characteristics of the telescopes used, of the order of 1000s of pulsars can populate the GC if distributed in a region of width $\sim$1$\,$pc or more around Sgr A*. 
For instance, if the pulsars are assumed to exist within the inner central pc ($\sigma_{\rm R}=1\,$pc), $\rm{N_{psr}^{GC}}\lesssim1600$ ($\rm{N_{psr}^{GC}}\lesssim5300$ in the case of using $\mathcal{L}_{1}$). 
A scenario of particular interest is when the assumed pulsar population in the GC is very compact around Sgr A*, within a volume containing (circular) pulsar orbital periods $P_{\rm b}\lesssim100\,$yr around \sgra. This smaller volume encloses 
the most promising pulsars for gravity tests
with \sgra \citep{wk99, pl04, liu12}, and is given by a $\sigma_{\rm R}\approx0.02\,$pc, yielding $\rm{N_{psr}^{GC}}\lesssim50$ ($\rm{N_{psr}^{GC}}\lesssim160$ in the case of using $\mathcal{L}_{1}$). Our results therefore cannot set stringent constraints on the number of pulsars in the inner parsec of the galaxy, mainly due to our low sensitivity at large radial scales 
(constrained by the wider beams of IRAM 30m and LMT).
However, on smaller radial scales ($\sigma_{\rm R}\lesssim0.1\,$pc), phased ALMA offers better sensitivity. In these smaller scales, we are still compatible with a population of tens to hundreds of pulsars close to the central SMBH and beamed towards us.

We note however that ${\rm N}_{\rm psr}^{\rm GC}$ is subject to big uncertainties. One of the largest ones arise from assumption made on the properties of the pulsars that may exist in the GC, from which the detection fraction, $f_{\rm s}$, is derived. To illustrate the impact of just a different mean spectral index taken for the GC pulsar population, the upper limit for $\rm{N_{psr}^{GC}}$ for a population within 1$\,$pc around Sgr A* varies between $\rm{N_{psr}^{GC}}\lesssim190$ and $\rm{N_{psr}^{GC}}\lesssim2100$ ($\rm{N_{psr}^{GC}}\lesssim620$ and $\rm{N_{psr}^{GC}}\lesssim7100$ if $\mathcal{L}_{1}$) for the spectral index distributions $\left<\alpha_2\right>=-1.4\pm1.0$ and $\left<\alpha_3\right>=-1.8\pm0.2$, respectively. In the case of a compact population enclosed within 0.02$\,$pc around \sgra, $\rm{N_{psr}^{GC}}$ varies between $\rm{N_{psr}^{GC}}\lesssim10$ and $\rm{N_{psr}^{GC}}\lesssim75$ ($\rm{N_{psr}^{GC}}\lesssim40$ and $\rm{N_{psr}^{GC}}\lesssim250$ if $\mathcal{L}_{1}$), again for $\left<\alpha_2\right>$ and $\left<\alpha_3\right>$, respectively. The upper limits on $\rm{N_{psr}^{GC}}$ for the scenarios with mean spectral indices $\left<\alpha_2\right>$ and $\left<\alpha_3\right>$, for comparison, are also shown in Figure~\ref{fig:Npsr_GC_Combined}.


Finally, we remark that the upper limit $\rm{N_{psr}^{GC}}$ refers to pulsars active in radio and beamed towards us. The limits for total number of neutron stars in the assumed regions would be considerably greater depending on the assumed beaming fraction and ratio of radio-active neutron stars over all the existing ones. For instance, for a mean beaming fraction of 0.1 \citep{faucher06} and a ratio of radio-active over all neutrons stars of $10^{-2}–10^{-3}$ \citep{faucher06, keanekramer08, sartore10}, the upper limits for neutron stars would scale up by a factor $\mathcal{O}(10^3–10^4)$. Therefore, despite our non-detections, our results are consistent with a large population of neutron stars in the Galactic Center; and compatible with results from theoretical simulations and population synthesis \citep{alexander17, baum18, grav20_precS2, chen2023}.

\section{Future improvements}\label{sec:futureimp}

\subsection{Flux density sensitivity}

The pulsar search in this paper used EHT 2017 campaign data with a total bandwidth of $\Delta\nu\approx4\,$GHz. Later EHT observations, starting from 2018, began to record an instantaneous bandwidth of $\Delta\nu=8\,$GHz, offering a factor of $\sqrt{2}$ improvement in system sensitivity. Additionally, for single-dish telescopes, observing in the VLBI mode requires a notably larger overhead compared to those in stand-alone mode, as a result of the time spent on phase calibrators and the associated slewing time (see Table~\ref{tab:observations_summary}). For the same observing length, dedicated pulsar surveys would largely increase the fraction of time spent on \mbox{Sgr$\,$A*}, by possibly more than a factor of 2; this would improve the system sensitivity by roughly another factor of 1.5.

Furthermore, in this work we searched the dataset individually which were in fact collected from simultaneous observations at different telescopes. In principle, it is possible to increase the search sensitivity by coherently summing the data collected with the entire EHT array, forming tied array beams towards areas in the vicinity of \mbox{Sgr$\,$A*} \citep[see e.g.,][]{bjk+16}. This may also effectively mitigate some of the systematics present in individual telescope data because they should not correlate between stations. A downside is that a large number of beams needs to be formed, which requires substantial computing power. For instance, given the longest baseline of the EHT 2017 observation, an order of 20 million beams would be needed to cover an area of 0.1\,arcsecond around \mbox{Sgr$\,$A*}. These beams will also have to be contained by the smallest beam in the array, which is the synthetic beam of ALMA for the EHT 2017 campaign. This could be improved if multi-beaming capability of ALMA becomes available. 

To improve the feasibility of coherent beam forming of the EHT array, the beams can be directed towards compact objects that may be identified by imaging observations. Alternatively, incoherently adding the data and/or coherently summing only a subset of the array, may also be considered as an option to increase the system sensitivity. Table~\ref{tab:sumtel} summarizes the delivered equivalent diameters of the array under different ways of summing the telescope signals. For coherent addition, the summation of the entire array will deliver a sensitivity equivalent to a 111-m dish, while summing only the most sensitive four telescopes in the array will still yield an equivalent diameter of 107\,m, only 7\% less in sensitivity. 

The sky area where a coherent beam may be formed is restricted by the smallest beam size of the array, which is from one of the interferometers. If summing only the single-dish telescopes, the sky coverage can then be enlarged to the smallest beam size of the single dishes. This will give an equivalent diameter of 64\,m, approximately 60\% more sensitive than LMT, the largest single dish in the array. 
For incoherent summation, adding the data from the entire array will give an equivalent diameter of 81\,m. However, only a few percent difference is expected from summing only the top three sensitive telescopes, i.e., ALMA, LMT and the NOrthern Extended Millimeter Array (NOEMA)\footnote{NOEMA, a twelve 15-m antennas interferometer located in the French Alps, with phasing capabilities offering an equivalent single-dish diameter $D_{\rm s}\approx52\,$m, joined the EHT array in 2018.}. Similarly, incoherent summation of all single-dish telescopes in the array, will deliver an equivalent diameter of 52\,m, only 8\% improvement in sensitivity from the LMT. This is largely due to the significant difference in gain among the telescopes. 

Finally, the sensitivity of the LMT during the EHT 2017 campaign is far from its full capacity, mainly due to the under-illumination of its surface. It is expected that the LMT data from later EHT observations will deliver much higher sensitivity. 


\begin{table}
    \centering
    \caption{The equivalent diameter delivered by a variety of summation forms of the EHT array. The equivalent diameters of each individual telescope are obtained from \cite{eht19b}.  }\label{tab:sumtel}
    \begin{tabular}{ccc}
    \hline
    \hline
Telescope combination & Diameter (m)\\
ALMA & 74 \\ 
LMT & 50 \\
    \hline 
Coherent, ALMA+LMT+NOEMA+PV & 107 \\
Coherent, all single dishes & 64 \\
Coherent, all & 111 \\
\hline
Incoherent, ALMA+LMT+NOEMA & 80 \\
Incoherent, all single dishes & 52 \\
Incoherent, all & 81 \\
\hline
\end{tabular}
\end{table}


\subsection{Orbital parameters and searching algorithms}

As discussed in Section~\ref{ssec:sensi_orb}, the acceleration search performed in this work has optimal sensitivity for pulsars in a certain range of orbits. To cover more compact orbits with \mbox{Sgr$\,$A*} optimally, or where a fast MSP is involved, it is possible to simply increase the range of acceleration and jerk used in the search (at the expense of higher computational costs). 

Alternatively, new searching schemes could also be introduced to directly explore the space of Keplerian parameters \citep[e.g.,][]{bcb+22}, delivering optimal sensitivity to more types of orbits. Nonetheless, the challenge for these search strategies is the computing power required, considering that pulsar orbits in the high stellar density region of the GC are expected to be significantly eccentric. As such, the full set of five parameters is required to describe the orbit rather than only three for circular orbits.

Another possible approach to enable optimal sensitivity for more compact orbits, is to divide up the entire length of data into segments and to search each segment individually \citep{ncb+15,lde+21,etd+21}. Admittedly, this approach suffers from a loss in sensitivity due to shorter integration times in each segment, but it is more computationally feasible.


\subsection{Data cleaning and quality}\label{sec:disc_cleaning}

The impact of red noise is a major source of sensitivity loss in pulsar searches, particularly detrimental in high radio frequencies, where the instrumentation and opacity variations produce an excess of power in the low-frequency end of the Fourier spectrum. This effect combines with the fact that radio magnetars, typically found slowly spinning, are one of the main targets of pulsar surveys in the short millimeter range, due to their typically-flat spectrum. An optimal red noise reduction scheme can improve the sensitivity of our surveys, although the exact filtering and parameters required are non-trivial and depend on the data properties and the specific properties of the pulsars to be detected \citep{singh22}. Future surveys may need different passes with a suite of red noise filtering techniques to optimize the results.

Extra improvements in sensitivity to pulsars could potentially come from the reduction of the locally-generated signals that exist in the data. The SIS receivers themselves and the downconversion systems in the (sub)millimeter telescopes are complex, with many oscillators and mixers, and therefore can introduce external signals into the final data recorded to disk. 
These local signals are often highly periodic. This is generally not an issue for spectroscopic or continuum observations, where the scans are shorter in integration length and the data are averaged over a comparatively longer time as compared to the data analyzed here for the pulsar search\footnote{As an example, typical spectroscopic observations use a sampling time longer than $\sim100\,$ms, which ``dilutes'' many of the fast periodic signals detected in a microsecond-order sampled data.}. Nonetheless, in the case of using the data for a search for pulsars, such local periodic signals can have a substantial impact on the sensitivity. 

The detrimental effect comes mainly from large excesses of power in many Fourier bins, which then interfere with the searching algorithms by the extra candidates found. Although local interfering signals (both internal and external to the telescope system) are common in radio observations, they can be particularly harmful at very high radio frequencies, where the dispersion effect is very small or negligible\footnote{At low radio frequencies, the de-dispersion step in a pulsar searching algorithm smears the power of broadband local signals, decreasing their impact in the detection of celestial signals by allowing to discern what broadband signals have traveled though the ISM.}. 

\subsection{Other wavelengths}

In addition of the presented search at $\uplambda=1.3\,$mm ($\nu=228\,$GHz), efforts to survey the Galactic Center pulsars using modern receivers were done at both centimeter \citep[$\nu\simeq$4 to 20$\,$GHz]{mq10, etd+21, acc+22} and short millimeter wavelengths \citep[$3\gtrsim\uplambda\gtrsim2\,$mm, $\nu\simeq$80 to 160$\,$GHz]{tor21, lde+21}, with no discoveries of pulsars.


With risk of having still a strong scattering affecting the centimeter-wavelength observations, and with the large impact on sensitivity by the steep spectrum of pulsars at the short millimeter wavelengths, attempting new surveys with high sensitivity in the short centimeter and long millimeter-wavelength regime is compelling. Such frequency range (around $\sim10-50\,$GHz) would still diminish the scattering and dispersive effects of the ISM, while observing in a spectral window with stronger emission from the steep-spectrum pulsars. This range may contain the frequency ``sweet spot'' between scattering mitigation and sensitivity to pulsars near Sgr$\,$A* \citep{mq10,macq15, bower18}.




\section{Summary and Conclusions}
\label{sec:concs}
We carried out a search for pulsars and fast transients in the Galactic Center with the Event Horizon Telescope 2017 observations, the first search at an observing wavelength of $\uplambda=1.3\,$mm ($\nu=228\,$GHz). The search used data from phased ALMA, LMT, and the \mbox{IRAM$\,$30$\,$m} telescope, the most sensitive telescopes in the EHT array of 2017. Periodicity searches both in the Fourier domain and with a Fast-Folding-Algorithm were conducted to each individual dataset collected on three different nights, with acceleration and jerk search incorporated to cope with the potential orbital motion of pulsars around \mbox{Sgr$\,$A*} or a binary companion. Single-pulse searches were also performed to the whole dataset.


These searches did not detect any pulsars or transients. We estimated the search sensitivity both theoretically and in practice by injecting artificial signals into the real data. The practical sensitivities for fast-spinning pulsars ($P=1-10$\,ms) are approximately 0.02, 0.4, 1\,mJy for ALMA, LMT and the \mbox{IRAM$\,$30$\,$m} telescope, respectively, but are roughly an order of magnitude worse for slow pulsars ($P=1-10$\,s). In addition, we showed the possible pulsar orbits that can be detected with our searching scheme. We explored the detectability of the search towards a simulated pulsar population in the Galactic Center and around Sgr$\,$A*, concluding that the sensitivity of the observations is still low. For MSPs in particular, which may be a dominant population in the region, the search sensitivity is close to zero. The lack of discoveries is therefore not indicative that pulsars do not exist in the region; there could still be pulsars in the covered areas that simply emit below our detection thresholds. 
Finally, we discussed future improvements that can be introduced to optimize the usage of these and similar data, to improve the system sensitivity and, overall, the chances for the detection of pulsars that likely still hide in the vicinity of the supermassive black hole Sgr$\,$A*.


\section*{acknowledgments}

We are grateful to the anonymous referee for the review and suggestions to improve the manuscript. We thank the staff at the participating observatories and correlator centers that made possible the EHT 2017 observations. PT thanks Pablo Mellado and William Robertson for their support through several stages of the data reduction in the IRAM servers. RPE is funded by the Chinese Academy of Sciences President's International Fellowship Initiative, Grant No. 2021FSM0004. SMR is a CIFAR Fellow and is supported by the NSF Physics Frontiers Center awards 1430284 and 2020265. This work was supported by the European Research Council Synergy Grant ``BlackHoleCam: Imaging the Event Horizon of Black Holes'' (grant 610058).
This paper makes use of the following ALMA data: ADS/JAO.ALMA\#2016.1.01404.V. ALMA is a partnership of the European Southern Observatory (ESO; Europe, representing its member states), NSF, and National Institutes of Natural Sciences of Japan, together with National Research Council (Canada), Ministry of Science and Technology (MOST; Taiwan), Academia Sinica Institute of Astronomy and Astrophysics (ASIAA; Taiwan), and Korea Astronomy and Space Science Institute (KASI; Republic of Korea), in cooperation with the Republic of Chile. The Joint ALMA Observatory is operated by ESO, Associated Universities, Inc. (AUI)/NRAO, and the National Astronomical Observatory of Japan (NAOJ). The National Radio Astronomy Observatory is a facility of the National Science Foundation operated under cooperative agreement by Associated Universities, Inc. 
The LMT is a project operated by the Instituto Nacional de Astr\'ofisica, \'Optica, y Electr\'onica (Mexico) and the University of Massachusetts at Amherst (USA). 
This work is partly based on observations carried out with the IRAM 30-m telescope under project number 084-17.
The IRAM 30-m telescope on Pico Veleta, Spain is operated by IRAM and supported by CNRS (Centre National de
la Recherche Scientifique, France), MPG (Max-Planck-Gesellschaft, Germany) and IGN (Instituto Geogr\'{a}fico
Nacional, Spain).
This research has made use of NASA's Astrophysics Data System Bibliographic Services. 
Part of this research was carried out at the Jet Propulsion Laboratory, California Institute of Technology, under a contract with the National Aeronautics and Space Administration.
The Event Horizon Telescope Collaboration thanks the following
organizations and programs: the Academia Sinica; the Academy
of Finland (projects 274477, 284495, 312496, 315721); the Agencia Nacional de Investigaci\'{o}n 
y Desarrollo (ANID), Chile via NCN$19\_058$ (TITANs) and Fondecyt 1221421, the Alexander
von Humboldt Stiftung; an Alfred P. Sloan Research Fellowship;
Allegro, the European ALMA Regional Centre node in the Netherlands, the NL astronomy
research network NOVA and the astronomy institutes of the University of Amsterdam, Leiden University and Radboud University;
the ALMA North America Development Fund; the Astrophysics and High Energy Physics programme by MCIN (with funding from European Union NextGenerationEU, PRTR-C17I1); the Black Hole Initiative, which is funded by grants from the John Templeton Foundation and the Gordon and Betty Moore Foundation (although the opinions expressed in this work are those of the author(s) 
and do not necessarily reflect the views of these Foundations); the Brinson Foundation; 
Chandra DD7-18089X and TM6-17006X; the China Scholarship
Council; the China Postdoctoral Science Foundation fellowships (2020M671266, 2022M712084); Consejo Nacional de Ciencia y Tecnolog\'{\i}a (CONACYT,
Mexico, projects  U0004-246083, U0004-259839, F0003-272050, M0037-279006, F0003-281692,
104497, 275201, 263356);
the Consejer\'{i}a de Econom\'{i}a, Conocimiento, 
Empresas y Universidad 
of the Junta de Andaluc\'{i}a (grant P18-FR-1769), the Consejo Superior de Investigaciones 
Cient\'{i}ficas (grant 2019AEP112);
the Delaney Family via the Delaney Family John A.
Wheeler Chair at Perimeter Institute; Direcci\'{o}n General
de Asuntos del Personal Acad\'{e}mico-Universidad
Nacional Aut\'{o}noma de M\'{e}xico (DGAPA-UNAM,
projects IN112417 and IN112820); 
the Dutch Organization for Scientific Research (NWO) for VICI award (grant 639.043.513), grant OCENW.KLEIN.113 and the Dutch Black Hole Consortium (with project number NWA 1292.19.202) of the research programme the National Science Agenda; the Dutch National Supercomputers, Cartesius and Snellius  
(NWO Grant 2021.013); 
the EACOA Fellowship awarded by the East Asia Core
Observatories Association, which consists of the Academia Sinica Institute of Astronomy and
Astrophysics, the National Astronomical Observatory of Japan, Center for Astronomical Mega-Science,
Chinese Academy of Sciences, and the Korea Astronomy and Space Science Institute; 
the European Union Horizon 2020
research and innovation programme under grant agreements
RadioNet (No 730562) and 
M2FINDERS (No 101018682); the Horizon ERC Grants 2021 programme under grant agreement No. 101040021;
the Generalitat
Valenciana (grants APOSTD/2018/177 and  ASFAE/2022/018) and
GenT Program (project CIDEGENT/2018/021); MICINN Research Project PID2019-108995GB-C22;
the European Research Council for advanced grant `JETSET: Launching, propagation and 
emission of relativistic jets from binary mergers and across mass scales' (Grant No. 884631); 
the Institute for Advanced Study; the Istituto Nazionale di Fisica
Nucleare (INFN) sezione di Napoli, iniziative specifiche
TEONGRAV; 
the International Max Planck Research
School for Astronomy and Astrophysics at the
Universities of Bonn and Cologne; 
DFG research grant ``Jet physics on horizon scales and beyond'' (Grant No. FR 4069/2-1);
Joint Columbia/Flatiron Postdoctoral Fellowship, 
research at the Flatiron Institute is supported by the Simons Foundation; 
the Japan Ministry of Education, Culture, Sports, Science and Technology (MEXT; grant JPMXP1020200109); 
the Japan Society for the Promotion of Science (JSPS) Grant-in-Aid for JSPS
Research Fellowship (JP17J08829); the Joint Institute for Computational Fundamental Science, Japan; the Key Research
Program of Frontier Sciences, Chinese Academy of
Sciences (CAS, grants QYZDJ-SSW-SLH057, QYZDJSSW-SYS008, ZDBS-LY-SLH011); 
the Leverhulme Trust Early Career Research
Fellowship; the Max-Planck-Gesellschaft (MPG);
the Max Planck Partner Group of the MPG and the
CAS; the MEXT/JSPS KAKENHI (grants 18KK0090, JP21H01137,
JP18H03721, JP18K13594, 18K03709, JP19K14761, 18H01245, 25120007); the Malaysian Fundamental Research Grant Scheme (FRGS) FRGS/1/2019/STG02/UM/02/6; the MIT International Science
and Technology Initiatives (MISTI) Funds; 
the Ministry of Science and Technology (MOST) of Taiwan (103-2119-M-001-010-MY2, 105-2112-M-001-025-MY3, 105-2119-M-001-042, 106-2112-M-001-011, 106-2119-M-001-013, 106-2119-M-001-027, 106-2923-M-001-005, 107-2119-M-001-017, 107-2119-M-001-020, 107-2119-M-001-041, 107-2119-M-110-005, 107-2923-M-001-009, 108-2112-M-001-048, 108-2112-M-001-051, 108-2923-M-001-002, 109-2112-M-001-025, 109-2124-M-001-005, 109-2923-M-001-001, 110-2112-M-003-007-MY2, 110-2112-M-001-033, 110-2124-M-001-007, and 110-2923-M-001-001);
the Ministry of Education (MoE) of Taiwan Yushan Young Scholar Program;
the Physics Division, National Center for Theoretical Sciences of Taiwan;
the National Aeronautics and
Space Administration (NASA, Fermi Guest Investigator
grant 80NSSC20K1567, NASA Astrophysics Theory Program grant 80NSSC20K0527, NASA NuSTAR award 
80NSSC20K0645); 
NASA Hubble Fellowship 
grants HST-HF2-51431.001-A, HST-HF2-51482.001-A awarded 
by the Space Telescope Science Institute, which is operated by the Association of Universities for 
Research in Astronomy, Inc., for NASA, under contract NAS5-26555; 
the National Institute of Natural Sciences (NINS) of Japan; the National
Key Research and Development Program of China
(grant 2016YFA0400704, 2017YFA0402703, 2016YFA0400702); the National
Science Foundation (NSF, grants AST-0096454,
AST-0352953, AST-0521233, AST-0705062, AST-0905844, AST-0922984, AST-1126433, AST-1140030,
DGE-1144085, AST-1207704, AST-1207730, AST-1207752, MRI-1228509, OPP-1248097, AST-1310896, AST-1440254, 
AST-1555365, AST-1614868, AST-1615796, AST-1715061, AST-1716327,  AST-1716536, OISE-1743747, AST-1816420, AST-1935980, AST-2034306); 
NSF Astronomy and Astrophysics Postdoctoral Fellowship (AST-1903847); 
the Natural Science Foundation of China (grants 11650110427, 10625314, 11721303, 11725312, 11873028, 11933007, 11991052, 11991053, 12192220, 12192223); 
the Natural Sciences and Engineering Research Council of
Canada (NSERC, including a Discovery Grant and
the NSERC Alexander Graham Bell Canada Graduate
Scholarships-Doctoral Program); the National Youth
Thousand Talents Program of China; the National Research
Foundation of Korea (the Global PhD Fellowship
Grant: grants NRF-2015H1A2A1033752, the Korea Research Fellowship Program:
NRF-2015H1D3A1066561, Brain Pool Program: 2019H1D3A1A01102564, 
Basic Research Support Grant 2019R1F1A1059721, 2021R1A6A3A01086420, 2022R1C1C1005255); 
Netherlands Research School for Astronomy (NOVA) Virtual Institute of Accretion (VIA) postdoctoral fellowships; 
Onsala Space Observatory (OSO) national infrastructure, for the provisioning
of its facilities/observational support (OSO receives
funding through the Swedish Research Council under
grant 2017-00648);  the Perimeter Institute for Theoretical
Physics (research at Perimeter Institute is supported
by the Government of Canada through the Department
of Innovation, Science and Economic Development
and by the Province of Ontario through the
Ministry of Research, Innovation and Science); the Princeton Gravity Initiative; the Spanish Ministerio de Ciencia e Innovaci\'{o}n (grants PGC2018-098915-B-C21, AYA2016-80889-P,
PID2019-108995GB-C21, PID2020-117404GB-C21); 
the University of Pretoria for financial aid in the provision of the new 
Cluster Server nodes and SuperMicro (USA) for a SEEDING GRANT approved towards these 
nodes in 2020;
the Shanghai Pilot Program for Basic Research, Chinese Academy of Science, 
Shanghai Branch (JCYJ-SHFY-2021-013);
the State Agency for Research of the Spanish MCIU through
the ``Center of Excellence Severo Ochoa'' award for
the Instituto de Astrof\'{i}sica de Andaluc\'{i}a (SEV-2017-
0709); the Spinoza Prize SPI 78-409; the South African Research Chairs Initiative, through the 
South African Radio Astronomy Observatory (SARAO, grant ID 77948),  which is a facility of the National 
Research Foundation (NRF), an agency of the Department of Science and Innovation (DSI) of South Africa; 
the Toray Science Foundation; the Swedish Research Council (VR); 
the US Department
of Energy (USDOE) through the Los Alamos National
Laboratory (operated by Triad National Security,
LLC, for the National Nuclear Security Administration
of the USDOE (Contract 89233218CNA000001); and the YCAA Prize Postdoctoral Fellowship.
This research used resources of the Oak Ridge Leadership Computing Facility at the Oak Ridge National
Laboratory, which is supported by the Office of Science of the U.S. Department of Energy under Contract
No. DE-AC05-00OR22725. We also thank the Center for Computational Astrophysics, National Astronomical Observatory of Japan.
The computing cluster of Shanghai VLBI correlator supported by the Special Fund 
for Astronomy from the Ministry of Finance in China is acknowledged.
This work was partially supported by FAPESP (Funda\c{c}\~ao de Amparo \'a Pesquisa do Estado de S\~ao Paulo) under grant 2021/01183-8.
APEX is a collaboration between the
Max-Planck-Institut f{\"u}r Radioastronomie (Germany),
ESO, and the Onsala Space Observatory (Sweden).  
The SMT is operated by the Arizona
Radio Observatory, a part of the Steward Observatory
of the University of Arizona, with financial support of
operations from the State of Arizona and financial support
for instrumentation development from the NSF.
Support for SPT participation in the EHT is provided by the National Science Foundation through award OPP-1852617 
to the University of Chicago. Partial support is also 
provided by the Kavli Institute of Cosmological Physics at the University of Chicago. The SPT hydrogen maser was provided on loan from the GLT, courtesy of ASIAA.
The SMA is a joint project between the SAO and ASIAA
and is funded by the Smithsonian Institution and the
Academia Sinica. The JCMT is operated by the East
Asian Observatory on behalf of the NAOJ, ASIAA, and
KASI, as well as the Ministry of Finance of China, Chinese
Academy of Sciences, and the National Key Research and Development
Program (No. 2017YFA0402700) of China
and Natural Science Foundation of China grant 11873028.
Additional funding support for the JCMT is provided by the Science
and Technologies Facility Council (UK) and participating
universities in the UK and Canada.
We acknowledge the significance that
Maunakea, where the SMA and JCMT EHT stations
are located, has for the indigenous Hawaiian people.
The EHTC has
received generous donations of FPGA chips from Xilinx
Inc., under the Xilinx University Program. The EHTC
has benefited from technology shared under open-source
license by the Collaboration for Astronomy Signal Processing
and Electronics Research (CASPER). The EHT
project is grateful to T4Science and Microsemi for their
assistance with Hydrogen Masers. 
We gratefully acknowledge the support provided by the extended
staff of the ALMA, both from the inception of
the ALMA Phasing Project
through the observational
campaigns of 2017 and 2018.
We would like to thank
A. Deller and W. Brisken for EHT-specific support with
the use of DiFX. 



%

\vspace{5mm}
\facilities{ALMA, LMT, IRAM:30m}


\software{ \texttt{MPIvdif2psrfits}, {\sc presto} \citep{ransom2011_presto_ascl}, {\sc riptide} \citep{2020MNRAS.497.4654M}, {\sc numpy} \citep{numpy_paper}, {\sc scipy} \citep{scipy_paper}, {\sc matplotlib} \citep{matplotlib_paper}, {\sc tempo} \citep{tempo_ascl}, {\sc sigpyproc} \citep{sigproc_ascl} }


\bibliography{EHT17_PSRsearch,psrrefs,modrefs}{}
\bibliographystyle{aasjournal}





\appendix

\section{Details of data properties}\label{sec:appA}

\begin{figure*}
     \centering
     \subfloat{
         \centering
         \includegraphics[width=\textwidth]{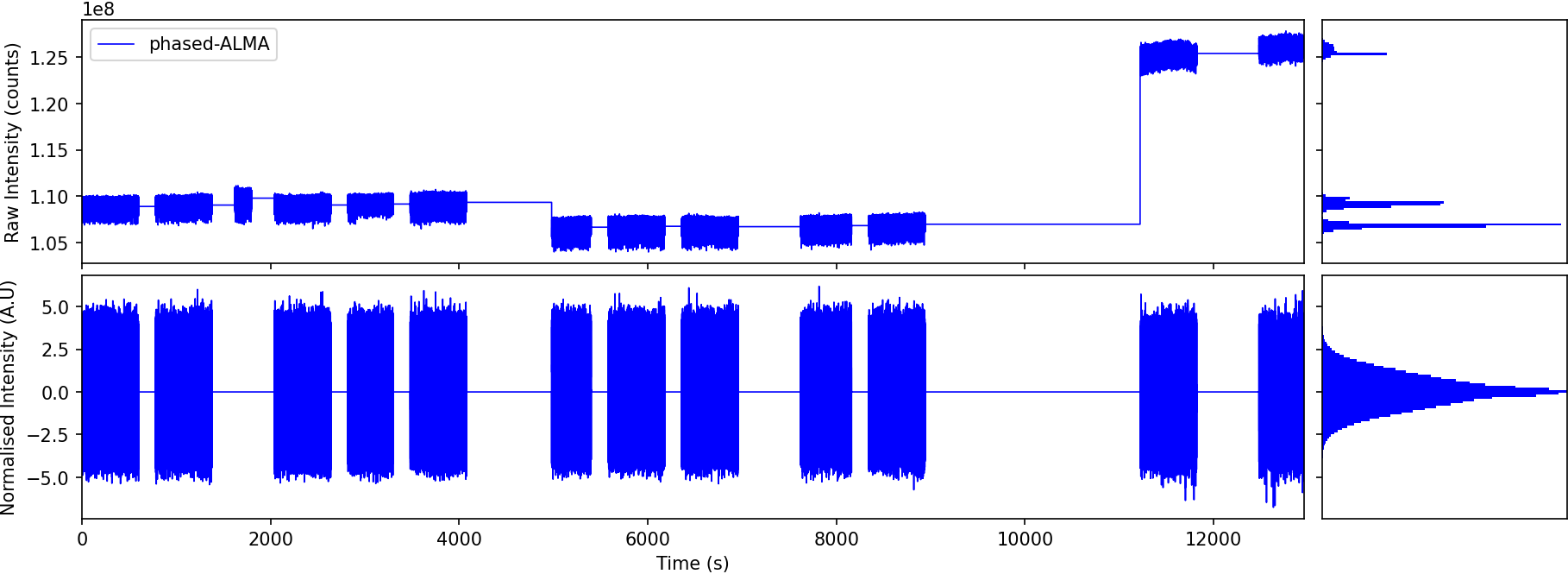}
     }\\
     \subfloat{
         \centering
         \includegraphics[width=\textwidth]{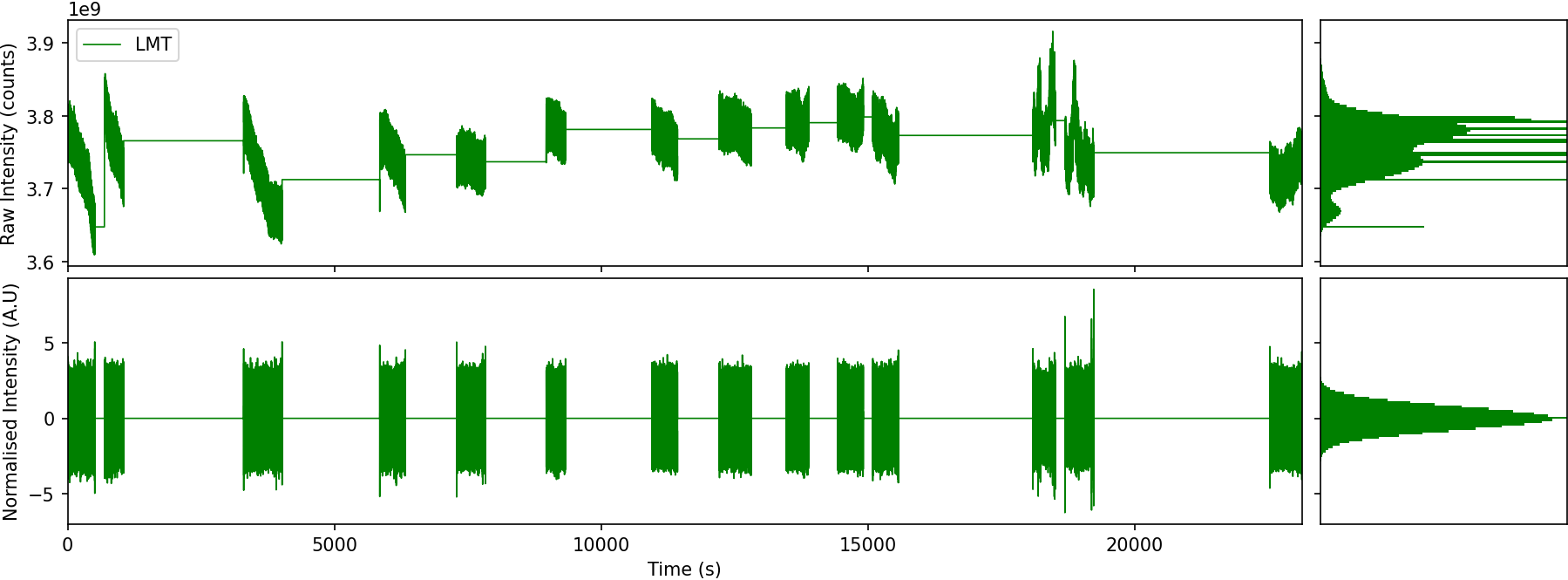}
    }\\
    \subfloat{
         \centering
         \includegraphics[width=\textwidth]{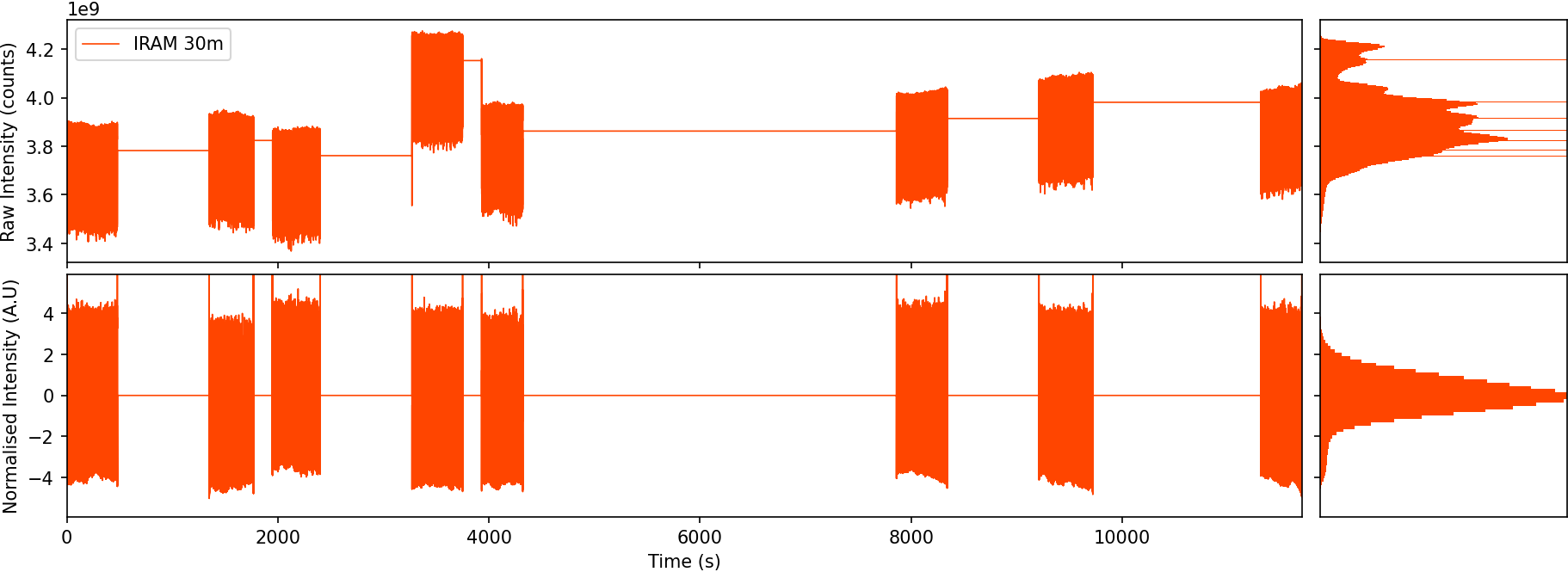}
     }
     \caption{Example time series of the observations of Sgr$\,$A* analyzed in this paper. The top two panels corresponds to data from phased-ALMA, the middle two panels to the LMT, and the bottom two panels to \mbox{IRAM$\,$30$\,$m} telescope data. Each pair of panels corresponding to one station show: (Top) the raw total intensity data just after the conversion to PSRFITS and (Bottom) the same total intensity time series after the preparation and cleaning described in Section~\ref{ssec:data_convprep}. A description of the main features of the data from each station is presented in Section~\ref{ssec:data_convprep}. The rightmost panels present a histogram of the distribution of the samples in linear scale. After the cleaning, the histograms tend to a Gaussian shape (although not perfect), indicating that the filtering and flagging schemes result in statistically better-behaved data. \label{fig:ts_examples}}
\end{figure*}

\begin{figure*}
     \centering
     \subfloat{
         \centering
         \includegraphics[width=\textwidth]{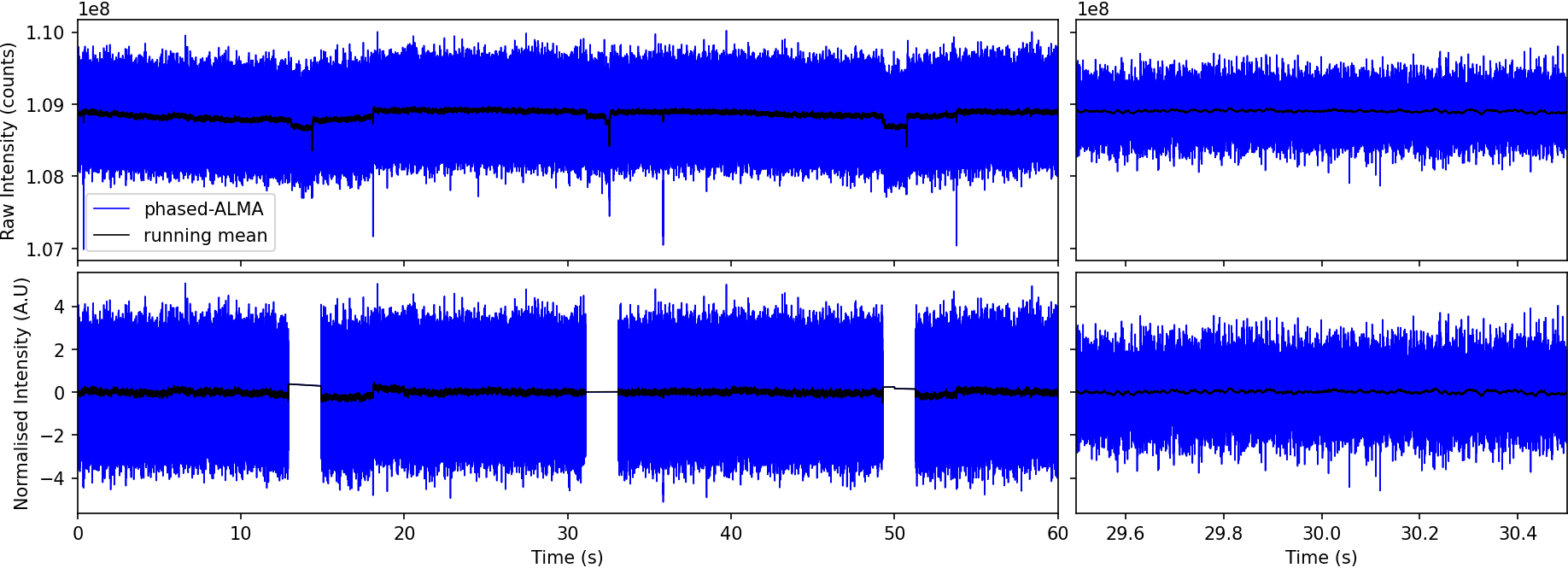}
     }\\
     \subfloat{
         \centering
         \includegraphics[width=\textwidth]{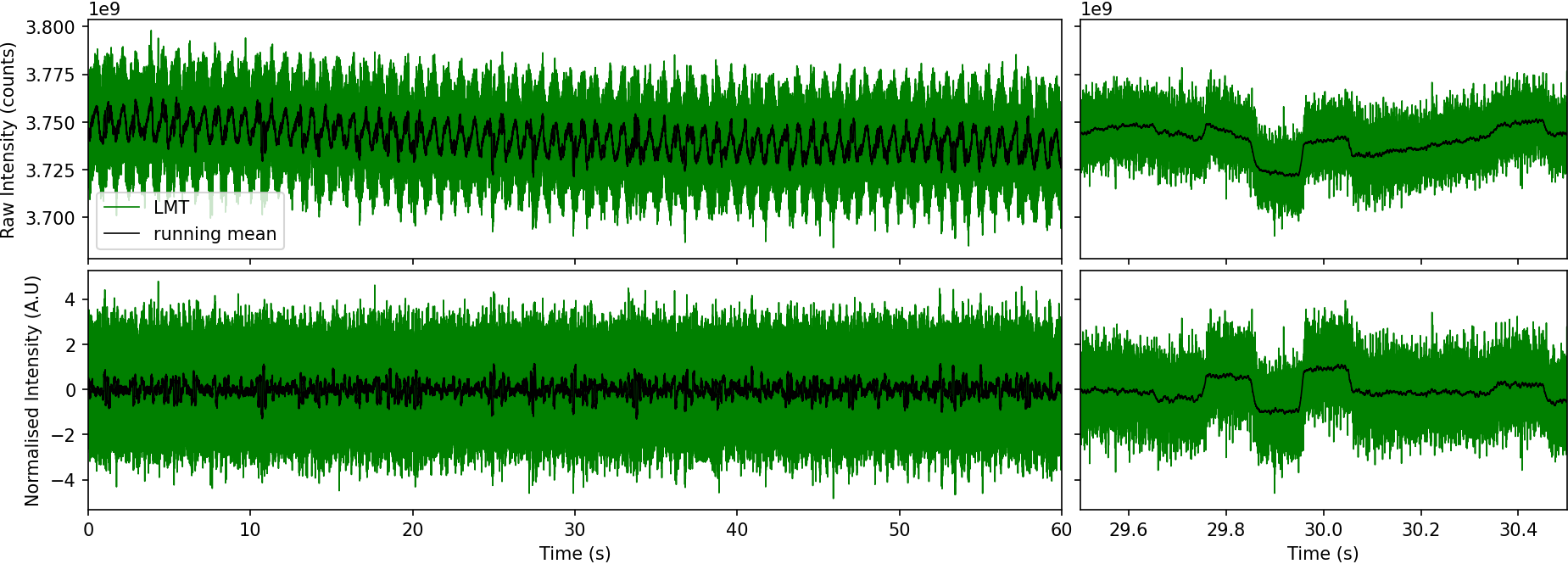}
     }\\
     \subfloat{
         \centering
         \includegraphics[width=\textwidth]{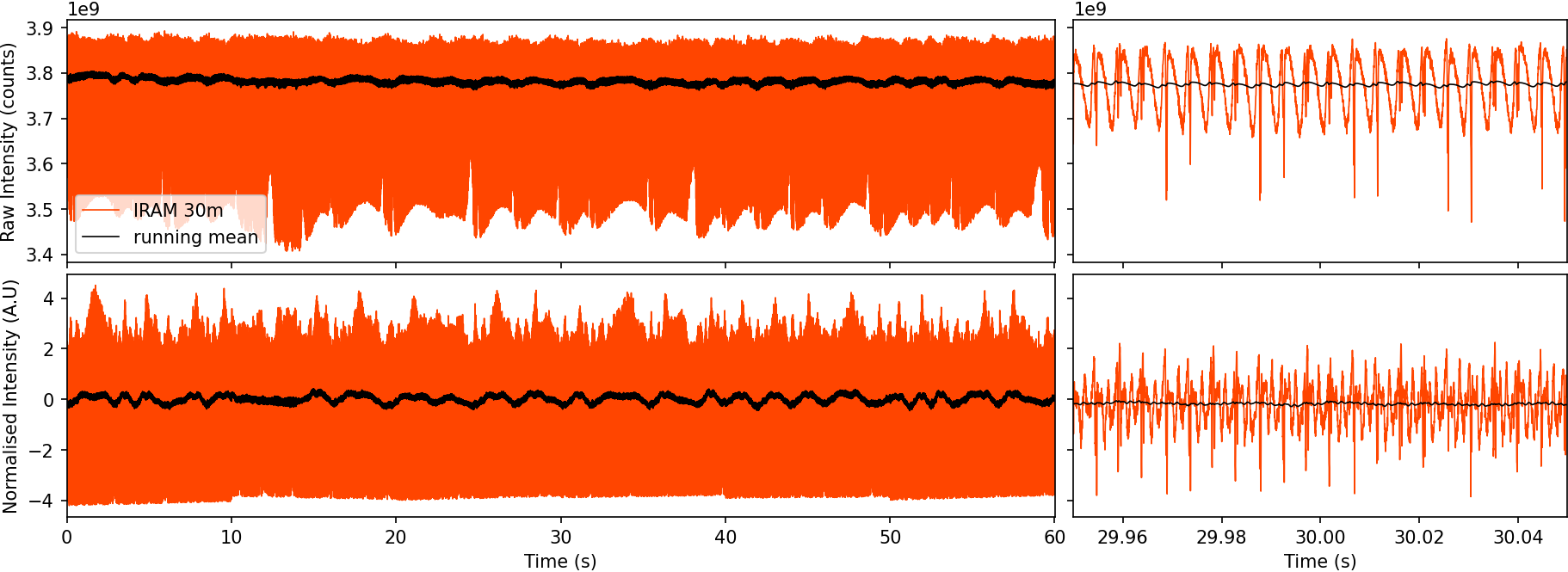}
     }
     \caption{Zoom to the example time series of the observations of Sgr$\,$A* analyzed in this paper. The top two panels corresponds to data from phased-ALMA, the middle two panels to the LMT, and the bottom two panels to \mbox{IRAM$\,$30$\,$m} telescope data. Each pair of panels corresponding to one station show: (Top) the raw total intensity data just after the conversion to PSRFITS and (Bottom) the same total intensity time series after the preparation and cleaning described in Section~\ref{ssec:data_convprep}. A description of the main features of the data from each station is presented in Section~\ref{ssec:data_convprep}. The panels on the right show an even closer zoom in, to facilitate the identification of the main issues of each dataset. Note that for the ALMA and LMT this extra zoom shows one second, while for the \mbox{IRAM$\,$30$\,$m} data, which suffer of much faster oscillations, the time axis of the right-hand panel encompasses only 0.1$\,$s. The black solid line shows a running mean with resolution of 10$\,$ms. \label{fig:ts_examples_ZOOM}}
\end{figure*}

In this Appendix we show example time series from the analyzed data of each of the three stations: phased ALMA, the LMT and the \mbox{IRAM$\,$30$\,$m} telescope. The data before and after the preparation and cleaning described in Section~{\ref{ssec:data_convprep}} are shown in Figures~\ref{fig:ts_examples} and \ref{fig:ts_examples_ZOOM}. Figure~\ref{fig:ts_examples} presents a broader view, with several scans and a relatively large part of each observation (not necessarily all of it). In Figure~\ref{fig:ts_examples_ZOOM} we show zoomed in versions, in which more details of the artifacts present in the data can be observed. Figure~\ref{fig:ts_examples_ZOOM} also shows that the artifacts are not completely removed after the cleaning steps, but at least the most prominent undesired signals are reduced. Despite the residual artifacts remaining in the data, the pre-processing tests injecting synthetic pulsar signals in the data show that our algorithms are capable of finding pulsars if they exist (with sufficient strength) in the data (see Section~\ref{sec:pipeverification}). Some sensitivity to pulsars is certainly lost due to the particular noise characteristics of the data. We try to model and account for these losses in our analysis (see Sections~\ref{sec:sys_sensi} and \ref{ssec:search_deepness}).

\end{CJK*}
\end{document}